\magnification=1200

\pageno=1
\centerline {\bf THE NONCRITICAL $W_\infty$ STRING SECTOR }
\centerline {\bf OF THE MEMBRANE   }
\medskip
\centerline { Carlos Castro }
\centerline { Physics Department}
\centerline { University of Texas}
\centerline { Austin , Texas 78757}
\centerline { World Laboratory, Lausanne, Switzerland }
\smallskip

\centerline {\bf December, 1996 }
\smallskip
\centerline {\bf ABSTRACT}

The  exact quantum integrability aspects of a $sector$ of  the membrane is investigated. It is found that spherical membranes ( in the lightcone gauge)   moving in flat target spacetime backgrounds admit a $class$ of integrable solutions linked to $SU(\infty)$
SDYM equations ( dimensionally reduced to one temporal dimension)  which, in turn, are  related to Plebanski $4D$ SD Gravitational equations. A further rotational Killing-symmetry reduction yields the $3D$ continuous Toda theory. It is precisely the latter which bears a direct relationship to non critical $W_\infty$ string theory. The expected critical dimensions for the ( super) membrane , ($D=11$) and $D=27$, are easily obtained.  This suggests that this  particular sector 
of the membrane's spectrum (connected to the 
$SU(\infty)$ SDYM equations ) bears a direct connection to a critical $W_\infty$ string spectrum $adjoined$ to a  $q=N+1$ unitary minimal model of the $W_N$ algebra in the $N\rightarrow \infty$ limit. Final comments are made about the connection 
 to Jevicki's observation that the $4D$ quantum membrane is linked to dilatonic-self dual 
gravity plus matter . $2D$ dilatonic ( super) gravity was studied by Ikeda and its relation to $nonlinear$ $W_\infty$ algebras from nonlinear integrable deformations of $4D$ self dual gravity was studied by the author.        
The full $SU(\infty)$ YM theory remains to be explored as well as the incipient role that noncritical nonlinear $W_\infty$ strings might have in the full quantization program.

\smallskip

\smallskip


 \centerline {\bf {1. Introduction }}

Recently,  exact solutions to $D=11$ spherical (super) membranes moving in flat target spacetime backgrounds were
constructed based on a particular class of  reductions of Yang-Mills equations from higher dimensions 
to four dimensions [1,2]. The starting point was dimensionally-reduced Super Yang-Mills theories based on the infinite dimensional
$SU(\infty)$ algebra. The latter algebra is isomorphic to the area-preserving diffeomorphisms of the sphere [3]. In this fashion the
super Toda molecule equation was recovered preserving one supersymmetry out of the $N=16$ expected. The expected critical target
spacetime dimensions for the ( super) membrane , $D=27 (11)$ , was closely related to that of an anomaly-free non-critical (super ) $W_{\infty}$
string theory. A BRST analysis revealed that the spectrum of the membrane should  have a relationship to the first unitary
minimal model of a $W_N$ algebra adjoined to a critical $W_N$ string in the $N\rightarrow \infty$ limit [1]. The class of particular solutions of the dimensionally-reduced $SU(\infty)$ YM equations studied are those of the type proposed by Ivanova and Popov  [2] which 
bears a direct relationship to $SU(\infty)$ instanton solutions in $4D$ that permits a  connection to the $SU(\infty)$ Toda molecule equation after  an specific ansatz is made [1].
 
The Toda theory emerges in other contexts $beyond$ the SDYM sector ( the self dual membrane ). It makes its appearance in noncritical $W_\infty$ strings; i.e.
$W_\infty$ gravity,  and in the quantum $4D$ membrane model studied by Jevicki 
[23]. A review of $SU(\infty)$ SDYM [1] is described in the next section
in connection to the Toda theory . In the final section we analyze in detail the role that noncritical $W_\infty$ strings have in the theory of membranes.
The critical dimensions, $D=27,11$ are recovered for the bosonic and supersymmetric case. Comments about the role that nonlinear noncritical $W_\infty$ strings in the full theory are made in the conclusion.

\centerline {\bf II}
\smallskip
\centerline {\bf 2.1 $SU(\infty)$ SDYM and the Toda Molecule }
\smallskip

Based on the observation that the spherical membrane (excluding the zero modes ) moving in $D$ spacetime dimensions, in the light-cone gauge, is essentially
equivalent to a $D-1$ Yang-Mills theory, dimensionally reduced to one time dimension, of the $SU(\infty)$ group ( see [8] for references); 
we look for solutions of the $D=10$ Yang-Mills equations (dimensionally-reduced to one temporal dimension). 
For an early  review on membranes  see Duff  [8] and the recent book by Ne'eman and Eizenberg [9].

Marquard et al [10] have shown that the light-cone gauge Lorentz algebra for the bosonic membrane is anomaly free iff $D=27$. The supermembrane critical dimension was found to be  $D=11$. To this date there is still some controversy 
about whether or not  the (super )  membrane is really anomaly free in these dimensions. They may suffer from other anomalies like $3D$ reparametrization invariance anomalies or global ones. What follows next  does $not$ depend on whether or not $D=27,D=11$ are  truly the  critical dimensions. What follows is just a straightforward quantization of a very special class of  solutions of  
the dimensionally-reduced  ( to one temporal dimension)  of $SU(\infty)$ YM equations, and which can be quantized exactly  due to  their  equivalence to the exactly integrable quantum continuous Toda molecule, obtained as a dimensional-reduction of the original continuous  $3D$ Toda theory [4,6] to $2D$ which is where $W_\infty$ strings live; this clarifies how a $3D$ membrane can have a connection to a $2D$ $W_\infty$ string.

We begin with the $D=10$ YM equations dimensionally reduced to one dimension. Let us focus on the bosonic sector of the theory. The supersymmetric case can be also analyzed via solutions to the supersymmetric Toda theory which has been discussed in detail in the literature . The particular class of solutions  one is interested in are those  of the type analyzed by Ivanova and Popov. Given :

$$\partial_a F_{ab} +[A_a,F_{ab}] =0.~A^\alpha_a T_\alpha \rightarrow A_a( x^b; q,p).~[ A_a,A_b] \rightarrow \{ A_a,A_b \}_{q,p}.
\eqno (2.1)$$
where the $SU(\infty)$ YM potentials [5] are obtained by replacing Lie-algebra valued potentials ( matrices) by $c$ number functions; Lie-algebra brackets   by
Poisson brackets w.r.t the two internal coordinates associated with the sphere; and the trace by an integral w.r.t these internal coordinates.  
In [1] we performed an ansatz following the results of Ivanova and Popov. The  $a,b,...   =8$ are   the transverse indices to the membrane after we performed the $10=2+8$ split of the original $D=10$ YM
equations. 

After the dimensional reduction to one dimension is done  we found that the following $D=10$ YM potentials, ${\cal A}$, -which will be later expressed in terms of the $D=4$ YM potentials, $A_1,A_2,A_3$ ($A_0$ can be gauged to zero )-  are
one class of solutions to the original $D=10$ equations iff they admit the following relationship :

$${\cal A}_1 =p_1 A_1.~{\cal A}_5 =p_2 A_1. ~{\cal A}_3 =p_1 A_3.{\cal A}_7 =p_2 A_3. \eqno (2.2a)$$
$${\cal A}_2 =p_1 A_2.~{\cal A}_6 =p_2 A_2.~{\cal A}_0 ={\cal A}_4 ={\cal A}_8 ={\cal A}_9 =0.  \eqno (2.2b)$$

where $p_1,p_2$ are constants and  $A_1,A_2,A_3$ are functions of $x_0,q,p$ only and obey the $SU(\infty)$ Nahm's equations :

$$\epsilon_{ijk}{\partial A_k \over \partial x_0} +\{A_i,A_j \}_{q,p} =0.~i,j,k =1,2,3. \eqno (2.3)$$

Nahm's equations are also obtained directly from reductions of $D=4$ Self Dual Yang-Mills equations to one dimension. The temporal variable $x_o
=p_1X_0+p_2X_4$ has a $correspondence $, not an identification, with the membrane's light-cone coordinate : $X^+=X^0+X^{10}$.  We refer to Ivanova and Popov and to our results in [1,2] for details.

Expanding $A_y =\sum A_{yl}(x_o)Y_{l,+1}.~A_{\bar y} =\sum A_{{\bar y}l}(x_o)Y_{l,-1}$; and $A_3$ in terms of $Y_{l0}$,
the ansatz which allows to recast the $SU(\infty)$ Nahm's equations as a Toda molecule equation is [1] :

$$\{A_y,A_{\bar y} \} =-i\sum^{\infty}_{l=1}~exp[K_{ll'}\theta_{l'}]Y_{l0} (\sigma_1,\sigma_2).~A_3
=-\sum^{\infty}_{l=1}{\partial \theta_l\over \partial \tau}.Y_{l0}. \eqno (2.4)$$ 

with $A_y ={A_1+iA_2\over \sqrt 2}.~A_{\bar y} ={A_1-iA_2\over \sqrt 2}.$

Hence, Nahm's equations become :

$$-{\partial^2 \theta_l\over \partial \tau^2 } =e^{K_{ll'} \theta_{l'}}.~~ l,l'=1,2,3....\eqno (2.5)$$

This is the $SU(N)$ Toda molecule equation in Minkowski form. The $\theta_l$ are the Toda fields where $SU(2)$ has been embedded
minimally into $SU(N)$. $K_{ll'}$ is the Cartan matrix which in the continuum limit becomes : $\delta''(t-t')$ [4]. The solution of the Toda theory is well known to the experts by now.  Solving for the  $\theta_l (\tau)$ fields and plugging their values into the first term of eq-(2.4) yields an infinite number of equations -in the $N\rightarrow \infty$ limit- for the infinite number of ``coefficients'' $A_{yl} (x_o),A_{{\bar y}l}(x_o)$. This allows to solve for the YM potentials $exactly$.  The ansatz [1] automatically yields the coefficients in the expansion of the $A_3$ component of the $SU(\infty)$ YM field given in the second term of (2.4). Upon quantization of the $SU(\infty)$ YM theory, the first term in eq-(2.4) is replaced by a commutator of two operators and as such the coefficients  $A_{yl} (x_o),A_{{\bar y}l}(x_o)$ become operators as well. The Toda fields become also operators in the Heisenberg representation. We will discuss the quantization of the Toda theory 
via the $W_\infty$ codajoint orbit method [25,26] below.

The
continuum limit of (2.5) is 

$$-{\partial^2 \theta (\tau,t) \over \partial \tau^2 } =exp~[\int~dt'\delta''(t-t') \theta (\tau,t')]. \eqno (2.6)$$

Or in alternative form :

$$-{\partial^2 \Psi(\tau,t) \over \partial \tau^2 } = \int~\delta'' (t-t')exp[\Psi (\tau,t')]~dt' ={\partial^2 e^{\Psi }\over
\partial t^2}. \eqno (2.7)$$

if one sets $K_{ll'}\theta_{l'} =\Psi_l$. The last two equations are the dimensional reduction of the $3D\rightarrow 2D$
continuous Toda equation given by Saveliev :

$${\partial^2 u(\tau,t)\over \partial \tau^2} =-{\partial^2 e^u\over \partial t^2}.~i\tau \equiv r=z+{\bar z}. \eqno (2.8a)$$
Eq-(2.8a) is referred as the $SU(\infty)$ Toda $molecule$ whereas 

$${\partial^2 u(z,{\bar z},t)\over \partial z \partial {\bar z}} =-{\partial^2 e^u\over \partial t^2}.\eqno (2.8b)$$
is the $3D$ continuous Toda equation which can 
obtained as rotational Killing symmetry reductions of Plebanski equations for Self-Dual Gravity in
$D=4$. Eq-(2.8a) is an effective $2D$ equation and in this fashion the original $3D$ membrane can be related to a $2D$ theory ( where the $W_\infty$ string lives in )  after the light-cone gauge is chosen. 

The Lagrangian ( and equations )for $4D$ SD gravity can be obtained from a dimensional reduction of the $SU(\infty)$ SDYM ( an effective six-dimensional one) [19,24] :

$${\cal L}= \int dz d {\tilde z}dy d{\tilde y}~{1\over 2}(\Theta_{,y}\Theta_{,z}
-\Theta_{,{\tilde y}}\Theta_{,{\tilde z}})+{1\over 3}\Theta\{\Theta_{,y},\Theta_{,{\tilde y}}\}.\eqno (2.9) $$
where $\Theta (z,{\tilde z},y,{\tilde y})$ is Plebanski's second heavenly 
form and the Poisson brackets are taken w.r.t $y,{\tilde y}$ variables.  A real slice can be taken by setting :  
${\tilde z}={\bar z},{\tilde y}={\bar y}.$

A rotatinal Killing symmetry reduction, $t\equiv y{\tilde y}$, 
yields the Lagrangian for the $3D$ Toda theory and a futher dimensional reduction $z+{\tilde z}=r$ gives the Toda molecule Lagrangian. 

From [19] we can find the explicit map between $\Theta$ and $\rho (r,t)$
$$A_y=\Theta_{,y};~A_{{\bar y}}=\Theta_{, {\bar y}};~A_z =-\Theta_{,{\bar y}}+f(y,{\bar y},z);~A_{{\bar z}} =\Theta_{,y}+g(y,{\bar y},{\bar z}).\eqno (2.10)$$
where $f,g$ are integration ``constants''. In the gauge $A_o=0\Rightarrow A_z=
A_{{\bar z}}$ and the two functions, $f,g$ are constrained to obey the latter  
condition plus the additional relations  stemming from the original 
 SDYM equations. Therefore, $f,g$ are fully determined. 

From our ansatz  (2.4) one can read off the correspondence between the Plebanski $\Theta$ ( after the corresponding reductions ) and the Toda field 
$\rho (r,t)$:

$$ 
\{\Theta_{,y},\Theta_{,{\bar  y}}\} \rightarrow 
{\partial^2 \over \partial t^2 }e^{\rho};
~{\partial \over \partial r}(-\Theta_{,{\bar y}}+f(y,{\bar y}))\rightarrow 
{\partial^2 \rho \over \partial r^2};~{\partial \over \partial r}A_z=
{\partial \over \partial r}A_{{\bar z}}.
\eqno (2.11a) $$

And, finally, one makes contact with Savaliev's Lagrangian of the Toda molecule :

$${\cal L}=\int dt~{1\over 2}({\partial^2 x\over \partial r \partial t})^2
+e^{(\partial^2 x/\partial t^2)}.~\rho (r,t)\equiv {\partial^2 x\over \partial t^2}. \eqno (2.11b)$$

Eqs-(2.9-2.11) are the essential equations that allows to extract the exact quantization of the Toda theory via the $W_\infty$ coadjoint orbit method described by [25].
Nissimov and Pacheva  have shown  that induced $W_\infty$ gravity could be seen as a WZNW model. They derived  the explicit form of the Wess-Zumino quantum effective action of chiral $W_\infty$ matter coupled to a chiral $W_\infty$ gravity background. The quantum effective action could be expressed as a geometric action on a coadjoint orbit of the deformed group of area-preserving diffs of the cylinder. A ``hidden'' $SL(\infty,R)$ Kac-Moody algebra was obtained as a consequence of the 
$SL(\infty,R)$ stationary subgroup of the $W_\infty$ codajoint orbit.  
 Yamagishi and Chapline earlier [26] proved  that an induced $4~D$ self-dual quantum gravity could be obtained via the $W_\infty$ coadjoint orbit method. An effective quantum action ( constructed in the twistor space)  was explicitly obtained as an infinite sum 
of two-dimensional effective Lagrangians with Polyakov two-dim lightcone gauge gravity as it fist term ( having a hidden $SL(2,R)$ Kac-Moody symmetry). The higher order terms are the result of the central extensions of the $W_\infty$ algebra.  

The crucial advantage  that the $W_\infty$ coadjoint orbit method has in the quantization of the continuous Toda theory is that an Operator Quantization method is extremely difficult. The ordinary Liouville theory , an $SL(2,R)$ Toda theory, is a notoriously difficult example. Its quantization using operator methods 
has taken years. Very recently, Fujiwara, Igarashi and Takimoto [21] have shown using exact operator solutions for quantum Liouville theory, based on canonical free field methods,  that the exact solutions proposed by Otto and Weight [22] are correct to all orders in the cosmological constant. They found that the hidden (quantum group) exhange algebra found by Gervais and Schnittger [20], 
$ U_q ( sl (2))$, was essential in order to maintain locality and  the 
operator form  of the field equation. In the continuous Toda case one expects a hidden quantum group $U_q (sl(\infty)$ structure. Not surprisingly, the appearance of the hidden $U_q (sl(\infty))$ must stem from the hidden $SL(\infty,R)$ Kac-Moody algebra associated with the stability subgroup of the $W_\infty$ coadjoint orbit method.     

The quantum effective action of the Toda theory is directly obtained from the exact results of [25,26] by the straightforward reduction given in eqs-(2.9-2.11) : one reads off the quantum effective action for the Toda theory directly from the ``dictionary'' between the $\Theta$ and the $\rho$ established in (2.9-2.11) $after$ performing the Darboux change of coordinates given by Plebanski.
 It is the latter   that expresses the second heavenly form,
 $\Theta$ in terms of the first heavenly form, $\Omega$. This is needed because the Chapline-Yamagishi  
quantum effective action is  given in terms of $\Omega$.  

Having discussed the importance of the Toda theory we proceed to study the role that noncritical $W_\infty$ strings have in the membrane quantization.

\centerline {\bf 2.2 A Membrane Sector  as a Non Critical $W_\infty$ String  }

\smallskip

In [1] we established the correspondence between the target space-times of non-critical $W_{\infty}$ strings and that of
membranes in $D=27$ dimensions. The supersymmetric case was also discussed and $D=11$ was retrieved. We shall review in further detail the construction [1].

The relevance of developing a $W_{\infty}$ conformal field theory ( with its quantum group extensions) has been emerging over the past years [28]. It was shown in [11] that
the effective induced action of $W_N$ gravity in the conformal gauge takes the form of a Toda action for the scalar fields and
the $W_N$ currents take the familiar free field form. The same action can be obtained from a constrained $WZNW$ model 
( modulo the global aspects of the theory due to  the topology. Tsutsui and Feher
have shown that richer structures emerge in the reduction process [29] ) 
by a quantum Drinfeld-Sokolov reduction process of the $SL(\infty,R)$ Kac-Moody algebra at the level $k$.   Each of
these quantum Toda actions posseses a $W_N$ symmetry. 

The authors [11] coupled $W_N$ matter to $W_N$ gravity in the conformal gauge, and integrating out the matter fields, they arrived at the induced effective action which was precisely the same as the Toda
action. It is not surprising that the Toda theory is related to $4~D$ Self-Dual gravity.  
In what follows, by $W_N$ string we mean the string associated to $WA_{N-1}$ algebra.

In general, non-critical $W_N$ strings are constructed the same way : by coupling 
$W_N$ matter to $W_N$ gravity. The matter and Liouville sector ( stemming from $W_N$ gravity) of the $W_N$ algebra can be
realized in terms of $N-1$ scalars, $\phi_k,\sigma_k$ repectively. These realizations in general have background charges which
are fixed by the Miura transformations 
[12,13]. The non-critical string is characterized by the central charges of the matter and
Liouville sectors, $c_m,c_L$. To achieve a nilpotent BRST operator these central charges must satisfy :

$$c_m+c_L =-c_{gh} =2\sum^N_{s=2} (6s^2-6s+1) =2(N-1)(2N^2+2N+1). \eqno (2.12)$$

In the $N\rightarrow \infty$ limit a zeta function regularization yields $c_m+c_L =-2$.

The authors [13] have shown that the BRST operator can be written as a sum of nilpotent BRST operators , $Q^n_N$, and that a nested basis can be chosen either for the Liouville sector or the matter sector but not for both. If the nested basis is chosen for the Liouville sector then [13] found that the central charge for the Liouville sector is :

$$c_L =(N-1)[1-2x^2N(N+1)]. \eqno (2.13)$$
were $x$ is an $arbitrary$ parameter which makes it possible to avoid the relation with the $W_N$ minimal models if one wishes to . Later we will show explicitly that 
the value of $c_L$ coincides precisely with the value of the central charge of a quantum Toda theory obtained from a quantum-Drinfeld-Sokolov reduction of the $SL(\infty,R)$ Kac-Moody algebra at the level $k $ such that $k+N =constant$ ( a constant that can be computed exactly)  in the $N\rightarrow \infty$ limit.  

By choosing  , if one wishes, $x$ appropriately one can, of course, get the $q^{th}$ unitary minimal models by fixing $x^2$ to be :
$$x^2_o =-2 -{1\over 2q(q+1)}. \eqno (2.14)$$
where $q$ is an integer. In this case, since $c_m +c_L=-c_{gh}$ , the central charge for the matter sector must be :
$$c_m =(N-1)(1-{N(N+1)\over q(q+1)}). \eqno (2.15)$$
which corresponds precisely  to the $q^{th}$ minimal model of the $W_N$ string as one intended to have by choosing the value of $x^2_o$. 
In the present case one has the freedom of selecting the minimal model since the value of $q$ is arbitrary. If $q=N$ then $c_m =0$ and the theory effectively reduces to that of the ``critical'' $W_N$ string. Conversely, if one chooses for the nested basis that corresponding to the matter sector instead of the Liouville sector, the roles of ``matter'' and ``Liouville'' are reversed. One would then have $c_L=0$ instead.

Noncritical strings involve two copies of the $W_N$ algebra. One for the matter sector and other for the Liouville sector. Since $W_N$ is nonlinear, one cannot add naively two realizations of it and obtain a third realization. Nevertheless there is a way in which this is possible [13]. This was achieved by using the nested sum of nilpotent BRST operators, $Q^n_N$. 
One requires to have all the matter fields , $\phi_k$; the scalars of the Liouville sector in the nested basis , $\sigma_{n-1},....\sigma_{N-1}$ plus the ghost and antighost fields of the spin $n,n+1,....N$ symmetries where $n$ ranges between $2$ and $N$. Central charges were computed for each set of the nested set of stress energy tensors, $T^n_N$ depending on all of the above fields which appear in the construction of the BRST charges : $Q^n_N$.

In order to find a spacetime interpretation, the coordinates $X^\mu$, must be related to a very specific scalar field of the Liouville sector ( since one decided to choose the nested basis in the Liouville sector) and that field is 
$\sigma^1$. 
It is this central charge, associated with the scalar  field $\sigma^1$,  that $always$ appears through its energy momentum tensor in the Miura basis.  
Because $\sigma _1$ always appears through its energy momentum tensor, it can be replaced  by an effective $T_{eff}$ of any conformal field theory as long as it has the same value of the central charge , $c=1+12\alpha^2\equiv 1-12x^2$, where $\alpha$ is a background charge.
$$T(\sigma ^1) =-{1\over 2}({\partial \sigma_1\over \partial z})^2 
-\alpha {\partial^2 \sigma_1 \over \partial z^2}. \eqno (2.16)$$
 
In particular, having $D$ worldsheet scalars, $X^\mu$, with a background charge vector, $\alpha_\mu$ :

$$T_{eff} =-{1\over 2}\partial_z X^\mu \partial_z X_\mu -\alpha_\mu.\partial^2_z X^\mu.~c_{eff} =D +12\alpha_\mu \alpha^\mu=1+12\alpha^2. \eqno (2.17)$$

For example, in the critical $W_\infty$ string case, one is bound to the unitary minimal models [12,13] and one must pick for central charge associated with the scalar, $\sigma_1$,  the value $\alpha^2=-x^2=-x^2_o$ given by (2.14).
The explicit value of $c$ of the critical $W_\infty$ string is obtained :
$$c_{crit}=1+12 (\alpha_o)^2 =1-12x^2_o = 1-12(-2-{1\over 2q(q+1)})= 25; q=N\rightarrow \infty. \eqno (2.18)$$

In the case of the ordinary critical string, $W_N=W_2$, $q=N=2$, one has  :

$$x^2_o =-2 -{1\over 2q(q+1)}\rightarrow -2-{1\over 12}\Rightarrow 
c_{eff}=1-12x^2_o =1+25 =26=c_{crit}. \eqno (2.19)$$

Since the parameter $x$ in the non-critical string case is an $arbitrary$ parameter that is no longer bound to be equal to $x_o$,  
the effective central charge in the non-critical $W_N$ string is now $1-12x^2$ in contradistinction to the critical $W_N$ string case : $1-12x^2_o$.  Therefore, if one wishes to make contact with $D=27~X^\mu$ scalars instead of $D=25$ 
one can choose $x$ in such 
a way that it obeys $1-12x^2 =(1-12x^2_o) +c_{m_o}$ where $c_{m_o}$ will turn out to be the central charge of the $q=N+1$ unitary minimal model of the $W_N$
algebra. Clearly, if one had chosen  $q=N$ instead, from eq-(2.15), one gets that $c_{m_o}\rightarrow 0$ and, as expected, the critical $W_N$ string is recovered : $x^2 \rightarrow x^2_o (q=N)$ given by (2.14).  If, in addition,         
one does $not$ wish to break the target space-time Lorentz invariance one 
$cannot$ have background charges for the $D~X^\mu$ coordinates. Therefore, for the case that $q=N+1 \Rightarrow c_{m_o}={2(N-1)\over N+1}$ ( instead of zero )  is obtained from  eq-(2.15), and the effective central charge is now  :

$$c_{eff}\equiv 1-12x^2 =(1-12x^2_o) +c_{m_o} = [26 -(1-{6\over (N+1)(N+2)})] +[2{N-1\over N+2}] \eqno (2.20)$$
then one concludes that 
$D=25+2=27=c_{eff}$ is recovered in the $N\rightarrow \infty$ limit. The reason why one wrote the last term of eq-(2.20) in such a peculiar way will be clarified shortly. 
In this way we have  shown  that the expected critical dimension for the  bosonic 
membrane  background , $D=27$,  has the same number of $X^\mu$ coordinates as that of a non-critical $W_{\infty}$ string 
background if one adjoins the $q=N+1$ unitary minimal model of the $W_{N}$ algebra to that of a critical $W_{N}$ string spectrum in
the $N\rightarrow \infty$ limit.
From eq-(2.20) one also  learns that 

$$D=2+25 =27 =1-12x^2 \Rightarrow 2x^2 =-{13\over 3}. \eqno (2.21)$$
which will be important to find the value of the central charge of the Toda theory, below. 

There are, of course, many other ways in which one could  recover
$D=27~X^\mu$ besides the way shown  in  eq-(2.20). The latter is a particular 
combination involving the critical $W_\infty$ string with the $q=N+1$
unitary minimal models. It is important to study the other possibilities.
Whatever these may be, these do not preclude the role that non-critical $W_\infty$ strings have in the theory.  The physical membrane spectrum has to contain a sector that should be related to a critical $W_\infty$ string adjoined to a $q=N+1$
unitary minimal model of the $W_{N}$ algebra in the $N\rightarrow \infty$ limit.
The full spectrum, moduli space of membrane vacua, etc...is far more vast than the slice furnished in (2.20). Our main point is that $W_\infty$ conformal field
theory, with its Kac-Moody extension, $W_\infty$ gravity,.... should contain 
important clues  to classify the spectrum and the moduli space of vacua, in the same way that ordinary conformal field theory did for the string. More precisely, we will argue that it is the non-linear extensions of the $W_\infty$ algebra  that must be involved if one wishes to relate to Jevicki's recent results [23].

The critical $W_\infty$ string   [12] is a  generalization of the ordinary string in the sense that instead of gauging the two-dimensional Virasoro algebra one gauges the higher conformal spin algebra generalization ; the $W_\infty$ algebra. The spectrum can be computed exactly and is equivalent to an infinite set of spectra of Virasoro strings with unusual central charges and intercepts [12]. As stated earlier , the critical 
$W_N$ string ( linked to the $A_{N-1}$ algebra) has for central charge the value ( $q=N$) :

$$c=1-12x^2_o= 26-(1-{6\over q(q+1)})=
26-(1-{6\over N(N+1)})=25\eqno (2.22)$$ 
where one has rewritten $25$ as $26-1$ to be able to make contact with the Virasoro unitary minimal models, as well,  given by the last term of eq-(2.20). This explains why the last term of (2.20) was written in such a peculiar way. 
Unitarity is achieved if the conformal-spin two-sector intercept is :

$$\omega_2 =1-{k^2-1\over 4N(N+1)}.~1\leq k\leq N-1. \eqno (2.23)$$

A particular  example of the above results is that in the ordinary non-critical ($W_2$) string there are many ways to 
have $c=26$. Choose for arbitrary value $x^2 =-2$ as opposed to non-arbitrary value of $x^2_o =-2-{1\over 12}$ required by  the $q=2$ Virasoro unitary minimal model. The central charge of the Liouville sector ( the nested basis) given by eq-(2.13) reads for $N=2$  :
$c_L=1(1+4.2.3)=25 =1-12x^2 =c_{eff}$, in this particular case the $c_L=c_{eff}$ and the $c_m =26-c_L =1$. And viceversa, choosing the matter sector to be in the nested basis, reverses  the roles of matter and Liouville, and one has 
$c_m =25;c_L =1$ which is the standard result that the ordinary 
$D=26$ critical string can be seen as a non-critical string in $D=25$ if one adjoins the Liouville mode that plays the role of the extra dimension.

It is not surprising in this picture of non-critical $W_\infty$ strings and 
quantization of $W_\infty$ gravity, to understand why there is the ubiquitous
presence of $W$ symmetry constraints in non-critical strings in $D\leq 1$ and matrix models in relation to the theory of integrable hierarchies. Paraphrasing [11] : ``the partition function of the (multi) matrix model, $Z$, which is related to the partition function of some low-dimensional string theory, or equivalently, two-dimensional gravity coupled to some $d\leq 1$ matter system, ${\cal Z}$, via the relation ${\cal Z}=log~Z$, is a $special$ solution of such integrable hierarchy. Special in the sense that it satisfies an extra constraint known as the string equation. In fact, $Z$ is itself subject to an $infinite$ number of constraints which form a Virasoro or $W$ algebra ``.  Furthermore, $W$ gravity coupled to $W$ matter is related to topological coset models  [11]. Whithin   our picture described here it is no surprise to see the appearance of $W$ symmetries in non-critical strings. One of the major advantages of $W$ conformal field theories is that allows the passage!
  of the string $c=1$ barrier.

An important remark is in order : we have to  emphasize that one should $not$ confuse $c_{eff}$ with $c_m,c_L$ in the same way that one must not confuse $x^2$ with $x^2_o$. The ordinary ($W_2$) string is a very $special$
 case insofar that $c_{eff}=c_m$ or  
$c_L$ depending on our choice for the nested basis.
The $D=27~X^\mu$ spacetime interpretation of the theory is $hidden$ in the stress energy tensor of the $\sigma^1$ field $T(\sigma_1) \rightarrow T(X^\mu)$ with 
$c_{eff}=c(D) =D=27$. And, in addition to the $27~X^\mu$, one still has the infinite number of scalars $\phi_1,\phi_2,....$ and the infinite number of remaining fields ,
$\sigma_2,\sigma_3,.....$ in the Liouville sector. Clearly the situation is vastly more complex that the string.  

From eqs-(2.12,2.13) one can infer that the value of the central charge of the matter sector, after a zeta function regularization, is $c_m =2+{1\over 24}$. The value of $c_L=-4-{1\over 24}$. And $c_{eff}=27$. The value of $c_m$ after regularization corresponds to the central charge of the first unitary minimal model of the $WA_{n-1}$ after $n$ is analytically continued to a negative value of 
$n =-146 \Rightarrow c(n)=2(n-1)/(n+2) =2+{1\over 24}$ [14]. The value of $c_L$ does not correspond to a minimal model but nevertheless corresponds to a very special value of $c$ where the $WA_{n-1}$ algebra truncates to that of the 
$W$ algebra associated with non-compact coset models [14] for specific values of the central charge :
$$WA_{n-1} \Rightarrow W(2,3,4,5) \sim { {\hat sl}(2,R)_n \over {\hat U}(1)}. \eqno (2.24)$$
this occurs at the value of $c(n) =2(1-2n)/(n-2) =-4-{1\over 24}$ for $n=146$.
This is another important clue that $W_\infty$ conformal field theory, with its Kac-Moody algebra extensions, rational and irrational, should reveal to us 
important information  of the membrane spectrum and its moduli  space of vacua.

The study of non-critical $W_\infty$ strings is very complicated in general. For example, ${\cal W} (2,3,4,5)$ strings are prohibitively 
complicated. One just needs to look into the cohomology of ordinary critical $W_{2,s}$ strings to realize this [12].
Nevertheless there is a way in which one can circumvent this problem when one restricts to the self dual solutions of the membrane. The answer lies in the integrability property of the
continuous Toda equation [4] . In the previous subsection we have shown how the exact quantization of the the Toda theory is automatically obtained by a straightforward dimensional reduction of the co-adjoint orbit quantization method  described by [25,26]. Furthermore, the quasi-finite highest weight irreducible representations of the  
$W_{1+\infty}, W_{\infty}$ algebras [30] allows to classify the co-adjoint orbits associated with these representations.  

We are going to proceed and calculate the value of the central charge of the Toda theory $without$ the need to quantize it explicitly! The quantum Toda theory has for central charge the value given in (2.13) for the specific value of $x^2$ found in eq-(2.21). i.e. if the BRST quantization of the continuous Toda action is devoid of $W_\infty$ anomalies the net central charge of the  matter plus Toda sector must equal $-2$ as we saw in eq-(2.12); i.e. $c_L=c_{Toda}$ ( after regularization). 

The central charge of the quantum $A_{N-1}$ Toda theory obtained from the quantum Drinfeld-Sokolov reduction of the ( noncompact version of $SU(\infty)$)  $SL(N,R)$ Kac-Moody algebra at level $k$ [11] is :

$$c_{Toda} =(N-1) (1-N(N+1) {(k+N-1)^2\over k+N}).~c_m+c_{toda}=-c_{gh}=c^{crit}. \eqno (2.25)$$
Another way of rewriting eq-(2.25) is from the Drinfeld-Sokolov reduction process :

$$c_{DS}=(N-1) -12 |\beta \rho -{1\over \beta}\rho^\nu |^2;~\beta ={1\over 
\sqrt {k+N}}. \eqno (2.26)$$
$\rho,\rho^\nu$ are the Weyl weight vectors of the $A_{N-1}$ Lie algebra and 
its  dual, respectively. One can read now the value of $x^2$ directly from 
eq-(2.13) and eq-(2.25) by equating $c_{Toda}=c_L$ :

$$2x^2=-{13\over 3}= (\beta -{1\over \beta})^2 =[{1\over \sqrt {k+N}}-\sqrt {k+N}]^2 ={ (k+N-1)^2 \over (k+N)}.\eqno (2.27)$$
The last equation allows us to compute explicitly the value of the coupling constant
appearing in the exponential function that gives the  interaction potential of the quantum Toda theory [6,15]. The Toda theory is conformally invariant and the conformally-
improved stress energy tensor obeys a Virasoro algebra with an adjustable central charge whose value depends on the coupling
constant of the exponential potential term appearing in the Toda action:
$ c(\beta)$.  
As pointed out in [11], it turns out that simply replacing the BRST operators by a normal-ordered version does not yield a nilpotent operator. In addition one has to allow for possible ( multiplicative) renormalizations of the stress-energy tensor appearing in the BRST charge. This is the origin of the 
$\beta =1/\sqrt{ k+N}$ factors. 

One may immediately notice that the expression for (2.27) is invariant under the exchange of $\beta \rightarrow 1/\beta $, the exchange of strong/weak coupling, does not alter the value of the central charge. This a good sign consistent with $S$ duality symmetry  of the alleged fundamental description of the membrane/string : $M,F,...$ theory.

One can now relate the value of the level ,$k$, of the $SL(N,R)$ Kac-Moody algebra   and $N$ in such a way that 
$k+N=\beta^{-2}$ is a  finite number when $N\rightarrow \infty$ ,   :

$$2x^2=(-13/3) =({1\over \sqrt {k+N}} -\sqrt {k+N})^2 = (\beta -{1\over \beta} )^2 \Rightarrow \beta^2 = {-7+ (-) \sqrt 13 \over 6}.\eqno (2.2.8)$$
The fact that $\beta =(k+N)^{-1/2}$ is purely imaginary should not concern us.
There exist integrable field theories known as Affine Toda theories whose coupling is imaginary but posseses soliton solutions
with real energy and momentum [15]. 
A natural choice is :  $k=-\infty$ so that $k+N=\beta^{-2}$ is $finite$  when $N\rightarrow \infty$.

The connection to the unitary Virasoro minimal models was established in 
eq-(2.22)( set $q=N+1$) :

$$D-2 =25 =c_{string} -[1-{6\over q(q+1)}] =26- [1-{6\over (N+1)(N+2)}]. 
\eqno (2.29)$$
This shall guide  us in repeating the arguments for the supersymmetric  case.  Similar arguments leads  to $D=11$ in the supermembrane case [1]. The argument proceeds as follows :  

Since $10$ is the critical dimension of the ordinary superstring the value of the central charge when one has $10$ worldsheet scalars  and fermions is $10(1+1/2)={30\over 2}$. In order to have the central charge of a critical super $W_\infty$ string one requires to have also the central charge of the super Virasoro
unitary minimal superconformal models  : $c_{Virasoro}=3/2$. 
The supersymmetric analog of the r.h.s of (2.29) is then  :

$$10(1+1/2) -c_{superconformal} ={30\over 2}-{3\over 2} ={27\over 2}. \eqno (2.30  )$$ 
The supersymmetric analog of the term $c_{m_o}={2(N-1)\over N+1}\rightarrow 2$, is : $2(1+1/2)=3$. One chooses the parameter $x^2$ in order to make contact with the bosonic sector of the $q=N+1$ unitary minimal model of the super $W_N$ algebra in the $N\rightarrow \infty$ limit.   
Writing down  the corresponding supersymmetric analog of each single one of the terms appearing in the r.h.s of eq-(2.20), and the same for the l.h.s , one has that
$D~X^\mu$ and $D~\psi^\mu$ ( anticommuting spacetime vectors and world sheet spinors) $without$ background charges yield a central charge $D(1+1/2)={3D\over 2}$; Therefore,  the supersymmetric extension of the corresponding terms of 
eq-(2.20) yields :

$${3D\over 2}=[10(1+1/2)-3/2]+[2(1+1/2)]=33/2 \Rightarrow D=11. \eqno (2.31)$$

Concluding , one obtains the expected critical dimensions for the (super) membrane if one $adjoins$ a $q=N+1$ unitary ( super) conformal minimal model of the ( super) $W_N$ algebra to a critical (super) $W_N$ string spectrum in the $N\rightarrow \infty$ limit. This all suggests that a sector of the  physical (super) membrane 
spectrum could be obtained exactly  the same way. Hence, in a heuristic manner,   we conjecture that :
there is a sector of the  physical ( super) membrane spectrum  that should be related to the  non-critical ( super) $W_\infty$ string constructed above. Furthermore, the quantum ( super ) membrane must be related to the quantization of (super) $W_\infty$ gravity. 
Further arguments that support our conjecture are given below.

What is required now is to quantize , upfront,  the membrane and to formulate the no-ghost theorem in order to confirm, if true, our conjecture. This is a very difficult problem. 
The full-fledge  membrane quantization  is a more arduous task. As explained in the previous subsection, the self dual sector is just the $SU(\infty)$ Self Dual Yang-Mills theory that can be related to the Toda theory after the dimensional reduction. In view of our findings about interpreting 
the membrane as a non-critical $W_\infty$ string with $c_{matter} +c_L =-2$ ( eq-(2.12)) and $c_L =c_{Toda}$, ( eqs-(2.13,2.25)),  suggests that a large sector of the $physical$  membrane spectrum, in addition to the one obtained by adjoining $q=N+1\rightarrow \infty$
unitary minimal $W_N$  conformal matter to a critical $W_\infty$ string,  might be obtained by adjoining  to the full quantized Toda ( or SDYM theory ) theory the remaining infinity of $W_\infty$ conformal scalar matter fields : $\phi_1,\phi_2,......\phi_k,....$ with the provision that the parameter $x$ is fixed by $c_{eff} (x)=1-12x^2 =27$ and
$c_m+c_L=-2$. ( Similar considerations apply to the supermembrane). 

Choosing a different value for $x$ and integer values for $c_{eff}$ suggests non-critical membrane backgrounds. An important diference between the ordinary critical string and  the critical ( super ) 
membrane is that the latter requires  an infinite number of fields in the  Toda sector ( a Liouville sector, $c_L$), an infinite number of fields in the  matter sector ( with central charge $c_m$) and the extra ( $D=11$) $D=27$ spacetime coordinates ( $c_{eff}$);
 in contradistinction to the critical string that only requires matter, $c_m =c_{eff}$ and no Liouville sector. This explains why $4D$ Self Dual Gravity ( which upon dimensional reduction yields the continuous Toda) must be a crucial player in the membrane quantization as pointed out by Jevicki in a different context [23].

Some time ago we were able to show that the $D=4$ $SU(\infty)$ ( super) SDYM equations  ( an effective $6$ dimensional theory) can be reduced to $4D$ ( super) Plebanski's  Self-Dual Gravitational equations  with spacetime signatures $(4,0);(2,2)$. The symmetry algebra of $D=4~SU(\infty)$ SDYM is a Kac-Moody extension of $W_\infty$ as shown recently by [31]. In particular, new hidden symmetries were found which are affine extensions of the Lorentz rotations. These new symmetries form a Kac-Moody-Virasoro type of algebra. By rotational Killing-symmetry reductions one obtains the $w_\infty$ algebra of the continuous Toda theory. For metrics with translational Killing symmetries one obtains the symmetry of the Gibbons-Hawking equations.

The relevance of 
Kac-Moody extensions of $W_\infty$ algebras has also been pointed out by Jevicki who has shown that the $4D$ membrane in the lightcone gauge yields a four dimensional world volume structure related to dilatonic-self dual gravity 
plus matter. The quantum theory is defined in terms of a $SU(\infty)$ Kac-Moody algebra. The quantization of $4D$ self dual gravity via the coadjoint orbit method has a hidden $SL(\infty,R)$ Kac-Moody algebra in the lightcone gauge. This is just the noncompact version of the $SU(\infty)$ Kac-Moody algebra. The presence of matter is also consistent with the presence of the matter fields $\phi_1,....\phi_k  $
with a central charge $c_m$. The matter terms [23] 
appear as a current-current interaction, $J^2$, where  the four-dimensional  field $J(x,t,\sigma^1,\sigma^2)$ found by Jevicki, originated  from a $SU(\infty)$ current algebra that accounts for the extra two-indices $\sigma^1,\sigma^2$.    

This is similar to what  we just found :  
$4D$ SD Gravity is a $reduction$ of $D=4~SU(\infty)~SDYM$ and the former is reduced to $\rightarrow$ the continuous $3D$ Toda theory. The Toda sector 
appears in noncritical $W_\infty$ strings  with the infinite number of scalars ( $W_\infty$ conformal  matter), $\phi_1,\phi_2,......$ . This construction must be related to the membrane's spectrum . What is left is the presence of the  sclar dilaton $\alpha (x,t)$ [23].  
$2D$ dilatonic supergravity was studied by Ikeda within the context of a $nonlinear$ gauge theory principle : one does not have a Lie bracket structure.  The nonlinear gauge principle allowed the author to construct a non-linear bracket that led to  $nonlinear$  $W_\infty$ algebras directly from nonlinear integrable perturbations of $4D$ self dual gravity. Hence, it seems that nonlinear ( but still integrable ) perturbations of self dual gravity span a richer sector than $W_\infty$ strings so it is very plausible that it is nonlinear noncritical $W_\infty$ strings that bear a closer connection to the full membrane. 
What is warranted is to establish, if possible, the relationship between the matter sector of Jevicki's Hamiltonian and ours.  
Integrable but linear deformations of self dual gravity were studied by Strachan [32]. The original $W_\infty$ algebra was constructed by [16]. The interpretation of such algebras as Moyal bracket deformations of the Poisson structure associated with the area-preserving diffs was found by [17,18].

This completes the review of [1]. We hope that we've clarified the interplay between $3D$ and $2D$ that appears after the light-cone gauge for the membrane is chosen and the importance that noncritical linear ( nonlinear) $W_\infty$ strings ; i.e. $W_\infty$ conformal field theory must have in the understanding of the membrane. Quantum Group extensions will come as no surprise. $W_\infty$ symmetries in string theory were also discussed by Zaikov [7].

\centerline {\bf Acknowledgements}

We thanks G. Sudarshan for advice . To Y. Ne'eman for having introduced me 
to membranes long before their current fashion; 
and to the World Laboratory, Lausanne, Switzerland for financial support.

\centerline {\bf References}

1. C. Castro : Journal  of Chaos, Solitons and Fractals. {\bf 7} 

no.5 (1996) 711. ``On the Exact Quantum Integrability of the Membrane ``

hep-th/9605029 ( under revision).

2. A. Ivanova, A.D. Popov : Jour. Math. Phys. {\bf 34} (1993) 674. 

3. J. Hoppe : "Quantum Theory of a Relativistic Surface" MIT Ph.D thesis (1982)

4. M.V. Saveliev : Theor. Math. Physics {\bf vol. 92}. no.3 (1992) 457.

5. E.G.F. Floratos, G.K. Leontaris : Phys. Lett {\bf B 223} (1989) 153

6. M. Toda : Phys. Reports {\bf 18} (1975) 1.

7. R. Zaikov : Phys. Letters {\bf B 211} (1988) 281.

Phys. Letters {\bf B 266} (1991) 303.

8. M.Duff : Class. Quant. Grav. {\bf 5} (1988) 189.

9 . Y. Ne'eman, E. Eizenberg : `` Membranes and Other Extendons `` World 

Scientific Lecture Notes in Physics. {\bf vol. 39} (1995).

10. U. Marquard, R. Kaiser, M. Scholl : Phys. Lett {\bf B 227} (1989) 234.

U. Marquard, M. Scholl : Phys. Lett {\bf B 227} (1989) 227.

11 .J. de Boer : "Extended Conformal Symmetry in Non-Critical String Theory" . 

Doctoral Thesis. University of Utrecht, Holland. (1993).

J. Goeree : `` Higher Spin Extensions of Two-Dimensional Gravity `` Doctoral 

Thesis, University of Utrecht, Holland, 1993.

J.de Boer, J. Goeree : Nucl. Phys. {\bf B 381} (1992) 329.

12. H.Lu, C.N. Pope, X.J. Wang: Int. J. Mod. Phys. Lett. {\bf A9} (1994) 1527. 

H.Lu, C.N. Pope, X.J. Wang, S.C. Zhao : "Critical and Non-Critical $W_{2,4}$ 

strings". CTP-TAMU-70-93. hepth-lanl-9311084.

H.Lu, C.N. Pope, K. Thielemans, X.J. Wang, S.C. Zhao : "Quantising Higher-Spin 

String Theories "  CTP-TAMU-24-94. hepth-lanl-9410005.

13. E. Bergshoeff, H. Boonstra, S. Panda, M. de Roo : Nucl. Phys. {\bf B 411} (

1994)  717.

 14. R.Blumenhagen, W. Eholzer, A. Honecker, K. Hornfeck, R. Hubel :" Unifying 

$W$ algebras".   Bonn-TH-94-01 April-94. hepth-lanl-9404113. 

Phys. Letts. {\bf B 332} ( 1994) 51

15 . M.A.C. Kneippe, D.I. Olive : Nucl.Phys. {\bf B 408} (1993) 565. 
  
Cambridge Univ. Press. (1989). Chapter 6,page 239.

 16. C.Pope, L.Romans, X. Shen : Phys. Lett. {\bf B 236} (1990)173. 

17. I.Bakas, B.Khesin, E.Kiritsis : Comm. Math. Phys. {\bf 151} (1993) 233.

18. D. Fairlie, J. Nuyts : Comm. Math. Phys. {\bf 134} (1990) 413.

 19. C.Castro :Phys. Lett {\bf 353 B} (1995) 201. Jour. Math. Phys. {\bf 34} (2

) (1993) 681

20. J.L. Gervais, J. Schnittger : Nucl. Phys {\bf B 431} (1994) 273.

 Nucl. Phys {\bf B 413} (1994) 277.

21. T. Fujiwara, H. Igarashi, Y. Takimoto : `` Quantum Exchange Algebra and Locality in Liouville Theory'' hep-th/9608040. 

22. H.J. Otto, G. Weight : Phys. Lett {\bf B 159} (1985) 341. 

Z.Phys. {\bf C 31} (1986) 219.

23. A. Jevicki :'' Matrix Models , Open Strings and Quantization of Membranes `` hep-th/ 9607187.

24. J. Plebanski, M. Przanowski :'' The Lagrangian of self dual gravitational field as a limit of the SDYM Lagrangian `` . hep-th/9605233. 

25. E. Nissimov, S. Pacheva : `` Induced $W_\infty$ Gravity as a WZNW Model ``

Ben Gurion University and Racah Institute  preprint : BGU-92/1/January-PH; RI-92  

26. K. Yamagishi, G. Chapline : Class. Quant. Gravity {\bf 8} (1991) 427.

27. N. Ikeda : `` $2D$ Gravity and Nonlinear Gauge Theory'' Kyoto-RIMS-953-93 preprint. 

28. P. Bouwnegt, K. Schouetens : Phys. Reports ${\bf 223}$ (1993) 183.

29. I. Tsutsui, L. Feher : Progress of Theor. Physics Suppl. {\bf 118} (1995)

173.

30. H. Awata, M. Fukuma, Y. Matsuo, S. Odake : Progress of Theor. Physics Suppl

. {\bf 118} (1995) 343.

31. A.D. Popov, M. Bordermann, H. Romer : `` Symmetries, Currents and Conservat

ion laws of Self Dual Gravity `` hep-th/9606077.

32. I. Strachan : Phys. Lett {\bf B 282} (1992) 63.

\bye
...........
.A.N.Leznov, M.V.Saveliev, I.A.Fedoseev : Sov. J. Part. Nucl. {\bf 16} no.1 (1

985) 81.

A.N.Leznov, M.V.Saveliev : "Group Theoretical Methods for Integration of Nonlin

ear Dynamical Systems " Nauka, Moscow, 1985.

. H.Awata, M.Fukuma, Y.Matsuo, S.Odake :"Representation Theory of the 

$W_{1+ {\infty}}$ Algebra".  RIMS-990 Kyoto preprint , Aug.1994.

S.Odake : Int.J.Mod.Phys. {\bf A7} no.25 (12) 6339.

. V.Kac, A. Radul : Comm. Math. Phys. {\bf 157} (1993) 429.

. M. Maggiore :''Black Holes as Quantum Membranes : A Path Integral Approach '' 

hepth-lanl-9404172. 

.J.L. Gervais, M.V. Saveliev : Nucl. Phys. {\bf B 453} (1995) 449.

. J.L. Gervais, Y. Matsuo : Comm. Mtah. Phys. {\bf 152} (1993) 317.

. A. Capelli, C.A.Trugenberger, G.R. Zemba :" Quantum Hall Fluids as 

$W_{1+\infty}$ Minimal Models"DFTT-9/95

. N. Berkovits, C.Vafa :Mod. Phys. Letters {\bf A9} (1994) 653.

. E. Guendelman, E. Nissimov, S. Pacheva . `` Volume-preserving diffeomorphis

ms versus Local Gauge Symmetry `` hep-th /950512 

A.Aurilia, A. Smailagic, E. Spallucci : Phys. Rev. {\bf D 47} (1993) 2536.

. C. Castro : ``p-branes as Composite Antisymmetric Tensor Field Theories ``

hep-th/9603117.

. D.B. Fairlie, J. Govaerts, A. Morozov : Nucl. Phys {\bf B 373} (1992) 214.

D.B. Fairlie, J. Govaerts : Phys. Lett {\bf B 281} (1992) 49.             .

-. B. Biran, E.G.F. Floratos, G.K. Saviddy : Phys. Lett {\bf B 198} (1987) 32

.

.S. Coleman : "Aspects of Symmetry " Selected Erice Lectures.  

. F. Yu, Y.S. Wu : Journal. Math. Phys {\bf 34} (1993) 5851-5895

.

. B. de Wit, J. Hoppe, H. Nicolai : Nucl. Phys. {\bf  B 305} (1988) 545.

B. de Wit, M. Luscher, H. Nicolai : Nucl. Phys. {\bf  B 320} (1989) 135.

.E.Bergshoeff, A.Salam, E.Sezgin ,Y.Tanii : Nucl. Phys. {\bf B 305} (1988) 

497.

E.Bergshoeff, E.Sezgin ,P.Townsend  :  Phys. Letts. {\bf B 189} (1987) 75.

 . N.Sanchez, H.de Vega : "String Theory in Cosmological Spacetimes" 

LPTHE-Paris-95-14

. M.Duff, R. Minasan :''Putting String/String Duality to the Test'' 

CTP-TAMU-16/94.  hepth-lanl-9406198.

. E.Witten : "String Theory Dynamics in Diverse Dimensions" IASSNS-HEP-95-18.

hepth/9503124

. M. Duff, R. Khuri, J.X. Lu : Phys. Reports {\bf 259} (1995) 213-326

. S. Kharchev, A. Mironov, A. Morozov : Theor. Math. Phys. {\bf 104} (1995)

866.

. K. Sfetsos : ``Nonabelian Duality, Parafermions and Supersymmetry `` hep-th

/9602179.

\bye

\bigskip
\centerline {\bf { 3 }}
\smallskip
\centerline {\bf 3.1 The SDYM Lagrangian and the Toda theory}
\smallskip

\centerline{\bf 3.2 The Exact Quantization of the Continuous Toda Field }
\centerline {\bf and the $W_\infty$ Coadjoint Orbit Method}
{\vbox {\vskip .2 truein}}   

Having reviewed the essential results of [1] permits us to look for classical solutions to the 
continous Toda equation and to implement the Quantization program presented in [5] after one takes the continuum limit which is what we are going to do in this section. 
The general solution to (2.8a) depending on two
variables , say $r\equiv z_+ +z_-$ and $t$ (not to be confused with time ) was given by [4]. 
The solution is determined by two
arbitrary functions , $\varphi (t)$ and $d(t)$. It is :

$$exp[-x(r,t)]=exp[-x_o(r,t)]\{1+\sum_{n> 1}~(-1)^n\sum_\omega\int
\int....exp[r\sum_{m=1}^n~\varphi (t_m)]~ \prod^{m=n}_{m=1}~dt_m d(t_m)$$
$$[\sum_{p=m}^n~\varphi
(t_p)]^{-1}[\sum_{q=m}^n~\varphi (t_{\omega (q)})]^{-1}. [\epsilon_m (\omega)\delta
(t-t_m)-\sum_{l=1}^{m-1}~\delta''(t_l-t_m)\theta [\omega^{-1}(m)-\omega^{-1}(l)]]\}. \eqno (3.1)$$
with :
$\rho_o=\partial^2x_o/\partial t^2 =r\varphi (t) +ln~d(t).$ This defines the boundary values of the solution 
$x(r,t)$ in the asymptotic region $r\rightarrow \infty $.
$\theta$ is the Heaviside step-function. 
$\omega$ is any permutation of the indices from $[2..........n] \rightarrow [j_2,..........j_n]$. 
$\omega (1)\equiv 1.$ $\epsilon_m (\omega)$ is a numerical coefficient. See [4] for details. 

An expansion of (3.1) yields :
$$exp[-x]=exp[-x_o]\{1-\mu +{1\over 2}\mu^2+........\}.\eqno (3.2)$$
where :
$$\mu \equiv {d(t)exp[r\varphi (t)]\over \varphi^2}. \eqno  (3.3)$$

The solution to the Quantum $A_{\infty}$  ( continous) Toda chain can be obtained by taking the continuum limit of 
the general solution to the finite nonperiodic Toda chain associated with the Lie algebra 
$A_N$ in the $N\rightarrow \infty$ limit. This is performed by taking the continuum limit of eqs-(30-34) and  eqs-(82-86) of [5] :

$$\varphi_i \rightarrow x_o(r,t).~\psi_{j_s} \rightarrow \partial^2 x_o/\partial t^2_s =r\varphi (t_s) +ln~d(t_s). \eqno (3.4)$$
In the $r\rightarrow \infty$ limit the latter tends to $r\varphi (t_s)$.
The continuum  limit of (86) in [5] is :

$$\sum_{j_1j_2...j_n} \rightarrow \int~\int~....dt_1 dt_2.......dt_n.~{\cal P}^1 \rightarrow [\sum \varphi (t_p) +O(\hbar )]^{-1}.
{\cal P}^2 \rightarrow [\sum \varphi (t_{\omega (q)}) +O(\hbar )]^{-1}.
\eqno (3.5)$$

Therefore, one just has to write down the quantum corrections 
to the  two factors $[\sum~\varphi]^{-1}$ appearing in  eq-(3.1) above. 
One must replace the first factor by a summation from $p=m$ to $p=n$ of terms like :
$$[\varphi (t_p) +O(\hbar)] \rightarrow \varphi (t_p)-{i \hbar \over w(t_p)} [{1\over w (t_p)}]_{,t_pt_p} -i \hbar\sum^n_{l=p+1} 
 {1\over w(t_l)} [{1\over w (t_l)}]_{,t_lt_l}. \eqno (3.6)$$ 

and the second factor  by  a summation from $q=m$ to $q=n$ of terms like :

$$[\varphi (t_{\omega (q)}) +O(\hbar )] \rightarrow [eq~(3.6) : ~p\rightarrow \omega (q)] +i \hbar \delta (t-t_{\omega (q)})$$
$$-i\hbar \sum_{l=1}^{q-1} [ w(t_{\omega (l)})]^{-1} [{1\over w(t_{\omega (l)})}]_{,t_{\omega (l)} t_{\omega (l)}} 
+i \hbar
\sum_{l=q+1}^n  [ w(t_{\omega (l)})]^{-1}     [{1\over w(t_{\omega (l)})}]_{,t_{\omega (l)} t_{\omega (l)}}    \eqno (3.7)$$

where $w (t)$ is a positive function that is the continuum limit of eqs-(30,34) of [5]. What one has done is to replace :

$${\hat k}_{j_mj_l}\equiv {{\tilde  k}_{j_mj_l}\over w_{j_l} w_{j_m}} \rightarrow \int~dt_m {\delta'' (t_m-t_l)\over w(t_l) w (t_m)} =
  {1\over w(t_l) }[{1\over w (t_l)}]_{,t_lt_l}\eqno (3.8a)$$
in all the equations in the continuum limit. One may  smear out the delta functions which appears in eq-(3.7) if one wishes so that  the denominators of eq- (3.1) are well defined.

These are the quantum corrections to the classical solution $\rho = \partial^2 x/\partial t^2 $ where $x(r,t )$ is given
in (3.1). These  are the continuum limits of eqs-(82-86) of [5].  It is important to realize that one must not add quantum
corrections to the $\varphi, d(t) $ appearing in the terms $exp[r\sum \varphi]$ and $x_o$ of (3.1). The former are two arbitrary
functions which parametrize the space of classical solutions. Upon quantization it follows from eq-(31) of [5] that $\varphi (t),d(t)$ become $r$ independent
( ``time'' independent ) operators obeying the equal $r$ ( time) commutation relations given by eq-(33) of [5] in the continuum limit  :
$$[\varphi (t),ln~d(t)]=-i{\hbar}{1\over w(t) }[{1\over w (t)}]_{,tt}.
\eqno (3.8b)$$
Therefore,  $\rho$ and $x$ acquire ${\hbar}$ quantum corrections given by 
(3.6,3.7) through the $c$-number function $w(t)$ terms and depend on  the non-commuting operators given by $\varphi (t), ln~d(t)$. Hence, $\rho$ or  $x$ should be seen as quantum operators acting on the Hilbert space of states associated with the quantization of the continuous Toda field : $\rho (r,t)$. Such states are
always  labeled as $|\rho>_{\varphi (t),d(t)}$. For convenience purposes we shall omit the suffix from now on but we should keep ithis in mind. 
Upon quantization, ${\hbar}$ appears and associated with Planck's constant a new parametric function has to appear : $w(t)$. One
has to incorporate also the coupling constant $\beta$ in all of the equations . This is achieved by rescaling the continuous
Cartan matrix by a factor of $\beta$ so that  $\partial^2 x/\partial t^2$ and  $\partial^2 x_o/\partial t^2$ are rescaled  by a
factor of $\beta$; i.e. $r\varphi$ acquires a factor of $\beta$ and $d(t)\rightarrow d(t)^\beta$.  
 Since $\beta$ is pure imaginary, for convergence purposes in the
$r=\infty$ region we must have that $\beta \varphi <0\Rightarrow \varphi =i\varphi$ also. In the rest of this section we will
work without the $\beta$ factors and only reinsert them at the end of the calculations. There is nothing unphysical about this
value of $\beta$ as we said earlier.

One of the integrals of motion is the energy. The continuos Toda chain is an exact integrable system in the sense 
that it posesses an infinite number of functionally independent integrals of motion : $I_n (p,\rho)$ in involution. i.e. 
The Poisson brackets  amongst $I_n,I_m$ is zero. Since these are integrals of motion, they do not depend on $r$. These
integrals can be evaluated most easily in the asymptotic region $r\rightarrow \infty$. This was performed in [2] for the case
that  $\varphi (t)$ was a negative real valued function which simplified the calculations. 
For this reason the energy
eigenvalue given in [2] must now be  rescaled by a factor of $\beta^2$  :  

$$E=\beta^2\int^{2\pi}_0~dt (\int^t~dt'\varphi (t'))^2. \eqno (3.9 
)$$

where we have chosen the range of the $t$ integration to be $[0,2\pi]$. Since $\beta\varphi <0 \Rightarrow \beta^2\varphi^2 >0$
and the energy is positive. We insist, once more, that $t$ is a parameter which is not the physical time and that $\varphi (t)$
does not acquire quantum corrections. The latter integral (22) is the eigenvalue of the Hamiltonian which is one of the Casimir
operators for the irreducible representations of $A_N$ in the $N\rightarrow \infty$ limit.  

The authors

found discrete energy levels for the quantum mechanical $SU(N)$ YM model. They did emphasize that discontinuities can occur in the $N=\infty$ limit. The reason that for finite $N$ a discrete spectrum occurs is due to the existence of a non-zero value of the Casimir energy. The classical membrane admits continuous deformations of zero area with no energy expense . Quantum mechanically, this classical instability is cured by quantum effects and any wave-function gets stuck in the potential valleys that become increasingly narrow the farther out one gets. Finite-energy wave functions fall-off rapidly and the quantum Hamiltonian is purely discrete. The ground state energy corresponds to the finite non-zero Casimir energy : the point beyond which no further deformations of the membrane into long stringlike configurations is possible.
In the supermembrane case the Casimir energy is zero and a continuous spectrum emerges. However, this was performed at a $finite$ value of $N$. Here we are 
discussing the different case ;  what happens at the $N=\infty$ limit.

\medskip
\centerline{\bf 4 The Spectrum Generating Algebra}
\smallskip

\centerline{\bf 4.1.  Highest Weight Representations}

We can borrow now the results by [6,7] on the quasi-finite highest weight irreducible representations of $W_{1+\infty} $
and $W_{\infty}$ algebras. The latter is a subalgebra of the former.  For each highest weight state, $|\lambda >$ parametrized by
a complex number $\lambda$ the authors [6,7] constructed  representations consisting of a finite number of states at each
energy level by succesive application of ladder-like operators. A suitable differential constraint on the generating function
$\Delta (x)$ for the highest weights $\Delta^\lambda_k$ of the representations was necessary in order to ensure that, indeed,
one has a finite number of states at each level. The highest weight states are defined :

$$W(z^n D^k) |\lambda> =0.~n\ge 1.k\ge 0.~~W(D^k) |\lambda>=\Delta^\lambda_k |\lambda > .~k\ge 0. \eqno (4.1)$$

The $W_{1+\infty}$ algebras can be defined as central extensions of the Lie algebra of differential operators on the circle.
$D\equiv zd/dz.~n\epsilon {\cal Z}$ and $k$ is a positive integer. The generators of the $W_{1+\infty}$ algebra are denoted by 
$W(z^n D^k)$; i.e. there is a one to one mapping between $z^n (z{\partial \over \partial z})^k$  and the $W_{1+\infty}$ generators. The $W_{\infty} $ generators  are obtained from the former : ${\tilde W} (z^n D^k) =W (z^n D^{k+1})$; where 
$(n,k \epsilon {\cal Z}.~k\ge 0)$. 
The commutation relations of the $W_\infty$ are :

$$[{\tilde W}[z^n f (z{\partial \over \partial z})],
{\tilde W}[z^m g(z{\partial \over \partial z})] ] = {\tilde W}[z^{n+m} f(z{\partial \over \partial z} +m)g(z {\partial \over \partial z})(z{\partial \over \partial z} +m)
- $$
$${\tilde W}[z^{n+m} f(z{\partial \over \partial z} )g(z {\partial \over \partial z}+n)(z{\partial \over \partial z} +n) +
C\Psi (z^n f(z{\partial \over \partial z})(z{\partial \over \partial z}),
z^m g(z{\partial \over \partial z})(z{\partial \over \partial z})). \eqno (4.2)$$

where $f,g$ are polynomials in $D\equiv z{\partial \over \partial z}$. The central charge term is given by the two-cocycle $\Psi$ times the constant $C$. ( see [6] for further references). The anti-chiral ${\bar W}_\infty$ is given exactly the same by replacing
everywhere $z\rightarrow {\bar z}$ and $\partial_z \rightarrow \partial_{{\bar z}}$. (There is no spin one current).

The generating function $\Delta (x)$ for the weights is :

$$\Delta (x) = \sum_{k=0}^{k=\infty} \Delta^\lambda_k~{x^k\over k!}. \eqno 
(4.3)$$
where we denoted explicitly the $\lambda$ dependence as a reminder that we are referring to the highest weight state $|\lambda
>$ and satisfies the differential equation required for quasi-finiteness :

$$b(d/dx)[(e^x-1)\Delta (x) +C] =0.~ b(w) =\Pi~(w-\lambda_i)^{m_i}.~\lambda_i\not= \lambda_j. \eqno (4.3)$$

$b(w)$ is the characteristic polynomial. $C$ is the central charge and the solution is :

$$\Delta (x) ={\sum_{i=1}^K~p_i(x)e^{\lambda_i x} -C\over e^x-1}.   \eqno (4.4)$$
The generating function for the $W_{\infty}$ case is ${\tilde \Delta} (x) =(d/dx) \Delta (x)$ and the central charge is 
$c=-2C$.

The Verma module is spanned by the states :

$$|v_\lambda> =W(z^{-n_1}D^{k_1})W(z^{-n_2}D^{k_2})...........W(z^{-n_m}D^{k_m})|\lambda >. \eqno (4.5)$$

The energy level is $\sum_{i=1}^{i=m}~n_i$. For further details we refer to 
[6,7]. Highest weight unitary representations for 
the $W_{\infty}$ algebra obtained from field realizations with central charge $c=2$ were constructed in [6].

The weights associated with the highest weight state $|\lambda >$ will be obtained from the expansion in (4.2).
In particular, the "energy" operator acting on $|\lambda >$ will be  :

$$W(D)|\lambda > =\Delta^\lambda_1 |\lambda >. \eqno (4.6)$$ 
$L_o = -W(D)$ counts the energy level :$[L_o, W(z^n D^k)] =-nW(z^n D^k)$.

As an example we can use for $\Delta (x)$ the one obtained in the free-field realization by free fermions or {\bf bc} ghosts 
[6]

$$ \Delta (x) =C{e^{\lambda x}-1\over e^x -1} \Rightarrow \partial \Delta/\partial \lambda =C{xe^{\lambda x}\over e^x -1}.
\eqno (4.7)$$
The second term is the generating function for the Bernoulli polynomials :

$${xe^{\lambda x}\over e^x -1} = 1+(\lambda -1/2)x +(\lambda^2-\lambda +1/6){x^2\over 2!} + (\lambda^3 -3/2
\lambda^2+1/2\lambda){x^3\over 3!} +.........\eqno (4.8)$$

Integrating (4.8) with respect to $\lambda $ yields back :

$$\Delta (x) =C{e^{\lambda x}-1\over e^x -1} =\sum_{k=0}~\Delta_k{x^k\over k!}. \eqno (4.9)$$

The first few weights (modulo a factor of $C$) are then :

$$\Delta_0= \lambda.~\Delta_1 = (1/2) (\lambda^2 -\lambda).~\Delta_2 = (1/3)\lambda^3 -(1/2)\lambda^2 +(1/6) \lambda.....\eqno
(4.10)$$

The generating function for the $W_{\infty}$ case is ${\tilde \Delta} (x) ={d\Delta (x)\over dx}\Rightarrow {\tilde
\Delta}^\lambda_k =\Delta^\lambda_{k+1}$.
This completes the short review of the results in [6,7]. Now we proceed to relate the construction of [6,7] with the results of section {\bf III}. 
\smallskip
\centerline{\bf 4.2  The $U_\infty$ Algebra}
\smallskip
We are going to construct explicitly the dimensional reduction of the 
$W_\infty \oplus {\bar W}_\infty $ algebra, the $U_\infty$ algebra, in terms of what one knows from the previous results in ${\bf 4.1}$.
From the previuos discussion we learnt that 
${\tilde \Delta}^\lambda_1=\Delta^\lambda_2$ is the weight associated with the "energy" operator. In the ordinary string, $W_2$
algebra, the Hamiltonian is related to the Virasoro generator, $H=L_o+{\bar L}_o$ and states are built in by applying the ladder-like
operators to the highest weight state, the "vacuum". In the $W_{1+\infty},W_{\infty}$ case it is $not$ longer true, as we shall see,  that the
Hamiltonian ( to be given later ) can be written exactly in terms of the zero modes w.r.t the $z,{\bar z}$ variables of the $W_2$ generator, once the realization of the
$W_{\infty}$ algebra is given  in terms of the $dressed $ continuous Toda field , $\Theta (z,{\bar z},t)$,  given by Savaliev [4].
The chiral generators are :

$$ {\tilde W}^+_2  =\int^{t_o}dt_1\int^{t_1}dt_2~exp[-\Theta (z,{\bar z};t_1)]{\partial\over \partial
z}exp[\Theta (z,{\bar z};t_1)-\Theta (z,{\bar z};t_2)] {\partial\over \partial z} exp [\Theta (z,{\bar z};t_2)]. \eqno (4.11a)$$

$$ {\tilde W}^+_n  =\int^{t_o}dt_1~\int^{t_1}dt_2....\int^{t_{n-1}}dt_n~
{\cal D}_+^{(0)}{\cal D}_+^{(1)}......{\cal D}_+^{(n-1)}exp[\Theta (z,{\bar z};t_n)].\eqno (4.11b) $$
with 

$${\cal D}^{(0)}_+ =
exp[-\Theta (z,{\bar z};t_1)]{\partial\over \partial
z};~ {\cal D}^{(m)}_+ \equiv exp[\Theta (z,{\bar z};t_m)-\Theta (z,{\bar z};
t_{m+1})] {\partial\over \partial z}.~m\ge 1.  \eqno (4.12)$$

The antichiral generators are obtained by replacing ${\partial \over \partial z}$ by ${\partial \over \partial {\bar z}}$ in eqs-(4.11,4.12). 
And now on, by continuous Toda field,  one means the $dressed$  continuous Toda field.

Hence, the chiral generators  have  the form $W^+_{h,0} [\partial^2 \rho/\partial z^2....\partial^h \rho /\partial z^h]$ [4]  
and the similar expression for the antichiral generators  $ W^-_{0,{\bar h}}$ is obtained by replacing $\partial_z\rightarrow 
\partial_{{\bar z}}$. 
After a dimensional reduction from $D=3\rightarrow D=2$ is taken, $r=z+{\bar z}$, one has :

$${\tilde W}_2 (r,t_o) =\int^{t_o}~dt_1~\int^{t_1}~dt_2~exp[-\rho (r,t_1)]{\partial\over \partial r}exp[\rho (r,t_1)-\rho
(r,t_2)] {\partial\over \partial r} exp [\rho (r,t_2)]. \eqno (4.13a)$$
And similar procedure applies to eqs-(4.11b,4.12) :

$$
{\tilde W}^+_n  =\int^{t_o}~dt_1~\int^{t_1}~dt_2~....\int^{t_{n-1}}~dt_n~
{\cal D}_+^{(0)}{\cal D}_+^{(1)}......{\cal D}_+^{(n-1)}exp[\rho(r;t_n)].\eqno (4.13b) $$
with :

$${\cal D}^{(0)}_+ = exp[-\rho(r;t_1)]{\partial\over \partial r};
~ {\cal D}^{(m)}_+ \equiv exp[\rho(r;t_m)-\rho(r;t_{m+1})] {\partial\over \partial r}. \eqno (4.14)$$

.....................................

When $\rho
(r,t)$ is quantized ( assuming that it can be done succesfully ); eqs-(3.6,3.7) involve the operator, ${\hat \rho}(r,t)$, acting on a suitable Hilbert space of states, say
$|\rho>$, and in order to evaluate (4.12) one needs to perform the highly complicated Operator Product Expansion between the
operators ${\hat \rho} (r,t_1),{\hat \rho} (r,t_2)$. Since these are no longer free fields it is no longer trivial to compute
per example the operator products :

$${\partial \rho \over \partial r}.e^\rho.~~e^{\rho (r,t_1)}.e^{\rho(r,t_2)}.....\eqno (4.15)$$

Quantization deforms the classical
$w_{\infty}$ algebra into $W_{\infty}$ [29,30]. For a proof that the $W_{\infty}$ algebra is the Moyal bracket deformation of
the $w_{\infty}$ see [30]. Later  in [31] we were able to construct the non-linear ${\hat W}_{\infty}$ algebras from
non-linear integrable deformations of Self Dual Gravity in $D=4$. 

One should not confuse ordinary Moyal deformations with Quantum-Group deformations with a natural Hopf-bialgebra structure ( co-product). Quantum-Group types of $W_N$ algebras, $q$-$W_N$, have been constructed by Awata et al [  ].
 
Since the $w_{\infty}$ algebra has been effectively quantized the
expectation value of the ${\tilde W}_2$ operator in the $\hbar \rightarrow 0$ limit, is related to the ${\tilde W}_2
(classical)$  given by (4.12). i.e; the classical Poisson bracket  algebra is retrieved by taking single contractions in the Operator Product Expansion of the quantum algebra. 
One can evaluate all expressions in the $r=\infty$ limit ( and set $d(t) =1$ for convenience.
The expectation value in the classical limit $<\rho|{\hat W}_2 ({\hat \rho}) |\rho>(\varphi (t))$
gives in the $r=\infty$ limit, after the dimensional reduction and after using  the asymptotic limits :

$${\partial \rho \over \partial r} =\varphi.~~~{\partial^2 \rho \over \partial r^2} = {\partial^2 e^\rho \over \partial
t^2}\rightarrow 0.~r\rightarrow \infty\eqno (4.14)$$
the value :

$$lim_{r\rightarrow \infty }<\rho|{\hat W}_2|\rho> = \int^{t_o}dt_1 \varphi (t_1) \int^{t_o}dt_1 \varphi (t_1).
\eqno (4.15)$$ 

after the normalization condition is chosen :

$$<\rho'|\rho>=\delta (\rho'-\rho).~<\rho|\rho> =1               \eqno (4.16) $$
We notice that eq-(4.15) is  the same as the integrand (3.9); so integrating 
(4.15) with respect to $t_o$ yields the energy
as expected.

It is useful to recall the results from ordinary $2D$ conformal field theory : given the  holomorphic current generator of
two-dimensional conformal transformations, $T (z)=W_2(z)$, the mode expansion is :

$$W_2 (z) =\sum_m~W^m_2 z^{-m-2}\Rightarrow W^m_2 =\oint~{dz\over 2\pi i} z^{m+2-1}W_2 (z). \eqno (4.17)$$
the closed integration contour encloses the origin. When the closed contour surrounds
$z=\infty$. This requires performing the conformal map $z\rightarrow (1/z)$ and replacing :

$$z\rightarrow (1/z).~dz\rightarrow (-dz/z^2).~W_2(z) \rightarrow (-1/z^2)^2 W_2 (1/z) =W_2(z)+{c\over 12}S[z',z]  \eqno (4.18)$$
in the integrand. $S[z',z]$ is the Schwarzian derivative of $z'=1/z$ w.r.t the $z$ variable.

There is also a $1-1$ correspondence between local fields and states in the Hilbert space :

$$|\phi> \leftrightarrow lim_{z,{\bar z}\rightarrow 0} {\hat \phi} (z,{\bar z}) |0,0>. \eqno (4.19)$$

This is usually referred as the $|in>$ state. A conformal transformation $z\rightarrow 1/z; {\bar z} \rightarrow 1/{\bar z}$:
defines the $<out|$ state at $z=\infty$

$$<out| =lim_{z,{\bar z}\rightarrow 0} <0,0|{\hat \phi} (1/z,1/{\bar z})(-1/z^2)^h (-1/{\bar z}^2)^{\bar h}  . \eqno (4.20)$$

where $h,{\bar h}$ are the conformal weights of the field $\phi (z,{\bar z})$.

The analog of eqs-(4.20) is to consider the state parametrized by $\varphi (t),d(t)$ :
$$|\rho>_{\varphi (t),d(t)} =lim_{ r\rightarrow \infty}~| \rho (r,t)>\equiv |\rho (out)>.$$      
 $$  |\rho>_{-\varphi (t),d(t)} =lim_{ r\rightarrow -\infty}~| \rho (r,t)>\equiv |\rho (in)>.\eqno (4.21)$$
since the continuous Toda equation is symmetric under $r\rightarrow -r\Rightarrow \rho (-r,t)$ is also a solution and it's
obtained from (3.1) by setting $\varphi \rightarrow -\varphi$ to ensure convergence at $r\rightarrow -\infty$.
As we pointed out earlier, the state $|\rho>$ is parametrized in terms of 
$\varphi (t),d(t)$ and for this reason one should always write it as
$|\rho>_{\varphi (t),d(t)}$ . What is required now is to establish the correspondence (a functor) between the representation space realized
in terms of the continuous Toda field and that representation ( the Verma module) built from the highest weight $|\lambda>$ 

$$<\lambda| {\tilde W} (D) |\lambda>={\tilde \Delta}^\lambda_1 \equiv \Delta^\lambda_2  \leftrightarrow 
<\rho|{\hat W}_2[{\hat \rho}(r,t)|\rho>.\eqno (4.22)$$

If one were to extract the zero mode piece of the ${\tilde W}_2$ operator via a contour integral around the origin , then integrate w.r.t. the $t$ variable and , finally, to  evaluate the expectation value,  one would arrive at a trivial  result.  There is a subtlety due the dimensional reduction from $2+1\rightarrow 1+1$ dimensions. If one $naively$ wrote down the expression   
  :

$$ {\cal P}[{ \Delta}^\lambda_1 ,{\bar  \Delta}^{\bar \lambda}_1]  \leftrightarrow 
<W^{0,0}_2>=\int^{2\pi}_0 dt_o~
_{\varphi}<\rho|[\oint{ dz\over 2\pi i }\oint{d{\bar z}\over
2\pi i} {\hat W}_2 (\rho(z,{\bar z},t))] |\rho>_{\varphi}. \eqno (4.23)$$ 
where one needs to add also the weights associated with the antichiral ${\bar W}_{\infty}$ algebra and the  real-valued generator  $W_2$ depends  on $r=z+{\bar z}$ and $t$ only. The l.h.s of (4.23) involves a dimensional reduction process in the weight space of the $W_\infty \oplus {\bar W}_\infty$ algebra which shall be discussed shortly. 

One now would be very hard pressed to avoid a zero answer after the contour integrations are completed in the r.h.s of (4.23). Expanding in powers of $(z+{\bar z})^n$ for negative and positive $n$ will give a trivial zero answer for the zero-mode of the $W_2$ operator. If one instead  expanded in positive powers of $(1/z+1/{\bar z})^n$; i.e. lets imagine 
expanding the function $e^{1/z}e^{1/{\bar z}}$ containing an essential singularity at the origin , in suitable powers of $z,{\bar z}$, the terms containing 
$z^{-1}{\bar z}^{-1}$ are the ones giving the residue. But $1/z +1/{\bar z} =(z+{\bar z})/z{\bar z}=r/z{\bar z}$ 
and this would contradict the assumption that all quantitites depend solely on the combination of $r=z+{\bar z}$ and $t$. Setting $t=z{\bar z}$ is incorrect because that will constrain the original $2+1$ Toda theory : $t$ is a parameter that appears in continuum graded Lie Algebras and has $nothing$ to do with the 
$z,{\bar z}$ coordinates. It plays the role of an extra ( compact ) coordinate, say an angle variable [4] but should not be confused with the $z,{\bar z}$ coordinates.

The correct procedure is to evaluate the generalization of what is meant by  eq-(4.17). The contour integral  means evaluating quantities for fixed times, which in the language of the $z,{\bar z}$ coordinates, implies choosing circles of $fixed$ radius around the origin and  integrating  w.r.t the angular variable. Therefore, the conserved Noether charges ( the Virsoro generators in the string case ) are just the  integrals of the conserved currents at fixed contour-radius ( fixed-times ). The  equal-time spatial ``hypersurfaces'' are then  the circles of fixed radius. The expression to evaluate is no longer (4.23) but 
the expectation value, say w.r.t the `in'' state, of the zero modes of the quantity  :

$${\hat I}_2 [f]= 
 \int^{2\pi}_0 dt'~f(t')~lim_{r'\rightarrow -\infty}~{\hat W}_2[\rho (r',t')]. \eqno (4.24)$$
where $f(t)=\sum_n a_n cos (nt)+b_n sin (nt)$. 

Hence, by $zero$ modes one means 
those w.r.t the angle variable $t$ and $not$ w.r.t the $z{\bar z}$ variables. The zero-mode generator's  expectation value of eq-(4.24) w.r.t the $|in>$ state  is now given by the   $n=0$ component :
$${ \Delta}_1 = <{\hat I}_2^0>= 
_{-\varphi(t)}<\rho |[ \int^{2\pi}_0 dt'a_0~lim_{r'\rightarrow -\infty}~{\hat W}_2[{ \hat \rho} (r',t')]]|\rho>_{-\varphi (t)}. \eqno (4.25)$$

Due to the dimensional reduction, in the l.h.s of (4.25) one must take  a suitable real-valued linear combination of the weights of the chiral and anti-chiral algebras. The weights live in a vector space and , as such, must be combined as vectors. First of all one imposes the conditions :
$ ({ \Delta}^\lambda_k)^* ={\bar  \Delta}^{\bar \lambda}_k.~{\bar \lambda}=
(\lambda)^*$ on all of the anti-chiral highest weights ; then one performs the linear combination of weights like those appearing in (4.10) :

$$\Delta_0=(\lambda +\lambda^*);~\Delta_1={1\over 2}[(\lambda +\lambda^*)^2-
(\lambda +\lambda^*)];$$
$$~\Delta_2 ={1\over 3}(\lambda +\lambda^*)^3 -
{1\over 2}(\lambda +\lambda^*)^2 +{1\over 6}(\lambda +\lambda^*);........\eqno (  )$$
What one is adding as vectors are the $weights$ and $not$ the values of 
$\lambda , \lambda^*$. Eq-(   ) is a particular example of how one would obtain the weights associated with the  dimensional reduction of the direct sum of the chiral and anti-chiral 
$W$ algebras. The most general case requires having a polynomial in $(\lambda +\lambda^*)$ of the type $\Delta_n={1\over 2}\sum (a_n +a_n^*)(\lambda +\lambda^*)^n$. 
Eq- (   ) will be from now on the $example$ of the dimensional-reduction 
process that  yields  the corresponding weights of the representations of the $U_\infty$ algebra from the original $W_\infty,{\bar W}_\infty$ representations.

The explicit  infinitesimal transformation law for the ${\hat \rho} (r,t)$ operator  under the 
$W_2$ transformations is

$$\delta^{\epsilon (t)} _{W_2}~{\hat \rho }(r,t) = 
 \int^{2\pi}_0 dt'~\epsilon (t'){\hat W}_2[{\hat \rho} (r',t')].{\hat \rho}(r,t).         \eqno (4.27)$$
Similarly for the remaining generators :
$$\delta^{\epsilon (t)} _{W_3}~{\hat \rho }(r,t) = 
 \int^{2\pi}_0 dt'~\epsilon (t'){\hat W}_3[{\hat \rho} (r',t')].{\hat \rho}(r,t).         \eqno (4.27)$$

$$\delta^{\epsilon (t)} _{W_n}~{\hat \rho }(r,t) = 
 \int^{2\pi}_0 dt'~\epsilon (t'){\hat W}_n[{\hat \rho} (r',t')].{\hat \rho}(r,t).         \eqno (4.27)$$
where the generators ( after the dimensional reduction ), $W_2,W_3,....$  are  $explicitly$ given in eqs-(    )

Going back to eq-(  ), the $n^{th}$ mode component is: 

$${\hat I}^{(2)}_n [C_{r'}] = 
 \int^{2\pi}_0 dt'~[a_ncos(nt')+b_nsin(nt)'] {\hat W}_2[\rho (r',t')].\eqno (  )$$
Similarly for the other generators, $W_s$, :

$${\hat I}^{(s)}_n [C_{r'}] = 
 \int^{2\pi}_0 dt'~[a_ncos(nt')+b_nsin(nt')]{\hat W}_s[\rho (r',t')]. \eqno (  )$$
where a radial quantization has been imposed :  $C_{r'}$ stands for a circle of fixed radius ( fixed time) $r'$ for all values of the ``angles'' :  $t'$ obtained by mapping the cylinder determined by the $r,t$ variables, $-\infty \leq r\leq +\infty$ and $0\leq t\leq 2\pi$ into the new complex plane
 $Z=e^{-i (t+ir)};{\bar Z}=e^{i(t-ir)}$. ( $z,{\bar z}$ are $not$ the same as 
$Z,{\bar Z}$).  

The explicit commutation relations of the $U_\infty$ algebra ( in the realization of eqs-(   )) require the knowledge of the OPE of the ${\hat W}_s$ generators and for this it is essential to have quantized the continuous Toda theory; i.e. perform the OPE of the exponentials of the Toda fields in eqs-(  ).  Shortly we will give the explicit commutation relations of $U_\infty$ in a different realization than the Toda one and which avoids the exact knowledge of the 
quantization program of the Toda theory.  If one has the Toda field realization, for example, the commutator  
$
[{\hat I}^{(s)}_m [C_{r_1}],{\hat I}^{(s')}_n [C_{r_2}]]$ can be computed by performing the OPE of terms like :
$$ 
\int^{2\pi}_0 dt_1~[a_mcos(mt_1)+b_msin(mt_1)]{\hat W}_s[\rho (r_1,t_1)$$
$$ 
\int^{2\pi}_0 dt_2~[a_ncos(nt_2)+b_nsin(nt_2)]~{\hat W}_s'[\rho (r_2,t_2)]. 
\eqno (  )$$
In the next subsection {\bf 4.3} we shall use our knowledge of conformal field theory methods to rewrite (   ) in a different form.  For example, in the Virasoro algebra case :

$$L_n =\oint {dz'\over 2\pi i}z'^{n+1} T(z'). 
~L_m =\oint {dz\over 2\pi i}z^{n+1} T(z).             \eqno ( )$$
$$[L_n,L_m]=\oint {dz'\over 2\pi i}z'^{n+1}\oint {dz\over 2\pi i}z^{n+1}
T(z').T(z)   $$

$$T(z').T(z)={1\over 2}{c\over (z-z')^4}+{2\over (z-z')^2}T(z) +{1\over (z'-z)}\partial_{z} T(z). \eqno (  )$$
both of the initial  contour integrals in (   )  enclose the origin. After the operator product is taken , the contour is deformed analytically  in such a way that the final contour integral encloses $z'$ but $not$ the origin. 
In this fashion one recovers the standard Virasoro algebra.     
Performing a change of variables $Z=e^{-i (t+ir)};{\bar Z}=e^{i(t-ir)}$ will allow us to write down the $U_\infty$ commutation relations as contour integrals in the $Z,{\bar Z}$ plane which is $not$ the same as the $z,{\bar z}$ complex plane.

It is possible to write down the commutation relations of the original $W_\infty$ algebra and its chiral counterpart, in a basis independent realization. It was proven by [      ] that the $W_\infty$ algebra could be obtained from the classical 
$w_\infty$ algebra by using the Moyal bracket deformations of the Lie-Poisson
bracket algebra ( after a suitable change of basis).  Adding cocycles allowed [  ] to generate the 
central extensions as well. The algebra [   ] must $not$ be confused with $q$-$W_N$ algebras discussed by [   ]. 

 The original Lie-Poisson bracket  inherent to $w_\infty$
algebras/ area-preserving diffeomorphisms can be defined in a basis-independent way as follows    
$[{\cal L}_f,{\cal L}_g]={\cal L}_{\{f,g\}}$.   
Where the infinite number of generators, ${\cal L}_f$ is parametrized by a family of functions, $f(q,p)$. The infinitesimal action of ${\cal L}_f$ on a function ,$F(q,p)$ is 

$$\delta^\epsilon_{{\cal L}_f} F=\{ \epsilon f,F\}.
~[{\cal L}_f,{\cal L}_g]={\cal L}_{\{f,g\}}\Rightarrow   
 [\delta^{\epsilon^1} _{{\cal L}_f},\delta^{\epsilon^2}_{{\cal L}_g}] F=
\{ \{\epsilon^1 f,\epsilon^2 g\}, F \}.
\eqno (  )$$
 
The bracket is the ordinary Poisson bracket between two arbitrary $f,g$ functions w.r.t the internal coordinates $q,p$ of the two-dimensional surface, a sphere, plane, cylinder,.... per example. The area-preserving diffs of the plane is the classical $w_\infty$ algebra. For a cylinder is the $w_{1+{\infty}}$ algebra  and for the sphere is $su(\infty)$. Locally the algebras are isomorphic but $not$ globally. If a particular basis-set of functions [  ]
( to expand the $f(q,p)$ functions)  is chosen, like : $w^{(i)}_m=q^{m+i-1}y^{i-1}$. The $i$ labels the particular set whereas the $m$ labels the particular function within the particular  $i^{th}$ set. One obtains the algebra [  ] :

$$\{w^{(i)}_m, w^{(j)}_n \}=[(j-1)m-(i-1)n]w^{(i+j-2)}_{m+n}. \eqno (  )$$
The generators $w^{(2)}_m$ are the conformal-spin $2$  Virasoro generators $L_m$ and in general $w^{(i)}_m$ are the conformal spin $i$ generators of the $w_\infty$ algebra. 

Eqs-(  ) can be generalized to the case in where the Moyal bracket replaces the Poisson-bracket as it was shown by [   ]. It is in this fashion that the $W_\infty$ algebra could be reconstructed. In particular, the $W_\infty$ algebra coincides precisely with the infinite-dimensional Lie algebra of Weyl-symbols of (pseudo) differential operators on the circle $S^1$ . The correspondence between (pseudo) differential operators and symbols is :

$$X(\xi,z)=\sum \xi^kX_k (z)\leftrightarrow \sum X_k (z)(-i\partial_z)^k.~
[X,Y] \equiv XoY-YoX;$$
$$XoY\equiv X(\xi,z)exp~[\partial_\xi \partial_z]Y(\xi,z).
\eqno (  )$$
where the $\partial_\xi, \partial_z$ act on the left and right respectively. 
The same steps can be applied to the anti-chiral algebra counterpart, ${\bar W}_\infty$ by replacing $z \rightarrow {\bar z}$ and 
$\partial_z \rightarrow \partial_{{\bar z}}$ in eqs-(  ).   
Once again, dimensional reductions of the algebra (  ) 
will reproduce automatically the 
$U_\infty$ algebra in  a realization which $differs$ from the Toda one.
The dimensionally reduced $U_\infty$ algebra is defined directly from eq-(4.2) by setting $z, {\bar z}\rightarrow  r$ and $\partial_z,\partial_{{\bar z}}\rightarrow \partial_r$ :

$$[{\tilde W}[r^n f (r{\partial \over \partial r})],
{\tilde W}[r^m g(r{\partial \over \partial r})] ] = {\tilde W}[r^{n+m} f(r{\partial \over \partial r} +m)g(r {\partial \over \partial r})(r{\partial \over \partial r} +m)-$$
$${\tilde W}[r^{n+m} f(r{\partial \over \partial r} )g(r {\partial \over \partial r}+n)(r{\partial \over \partial r} +n) +
C\Psi (r^n f(r{\partial \over \partial r})(r{\partial \over \partial r}),
r^m g(r{\partial \over \partial r})(r{\partial \over \partial r})). \eqno (4.2)$$
without the need to quantize the Toda theory and to evaluate the OPE of the Toda field exponentials and imposing normal ordering conditions in the defining generators, etc...;

Now we present our main and only postulate. 

The spectrum generating  $U_\infty$ algebra of the Toda theory is a dimensional reduction of the direct sum of $W_\infty $ and ${\bar W}_\infty $ indicated by eqs- (   ). 
 Eventhough one cannot write down  explicitly the commutation relations (    )  until one has a grasp of the full quantization program of the continuous Toda theory with its OPE of the exponentials of the Toda field and its associated Quantum Group structures, Exchange Algebra...., one can still bypass the 
latter quantization process and : 
postulate that for every highest weight representation of the original $W_\infty, {\bar W}_\infty$ algebras constructed by [6] one can always write down the corresponding   weights explicitly ( after following the dimensional reduction procedure of eqs- 
( )) and plug these weights   into  the l.h.s of eqs-(   ).  The latter equations now 
can be used to $define$,  in an implicit and indirect manner,  the corresponding $|\rho>$ state. The converse is not true. For every quantum state
$|\rho>$, appearing in the r.h.s of (   ) one cannot claim that it always  
corresponds to a particular highest weight representation.

To recapitulate : The spectrum generating $U_\infty$ 
symmetry algebra acting on the Toda molecule, stems from the $Self~Dual~SU(\infty)$ 
Supersymmetric Gauge Quantum Mechanical Model associated with the light-cone gauge of the $Self Dual$ ( spherical)  supermembrane : a dimensionally-reduced super SDYM theory  to $one$ temporal dimension. The membrane's time coordinate ( $X^+$)  has a $correspondence $ with the   $r=z+{\bar z}$ variable. The extra coordinate arises from the $t$ parameter so the initial $3D$ continuos  ( super ) Toda theory 
is dim-reduced to  a $1+1$ ( super ) Toda $molecule$ : $\Theta (z,{\bar z},t)\rightarrow \rho (r,t)$ and, in this way,  an $effective$ two-dimensional theory emerges. Hence, the intrinsic $3D$ Self Dual membrane spectrum can be obtained from the $U_\infty$ symmetry algebra ( spectrum generating algebra ) of the 
effective two-dimensional ( super) Toda molecule theory.       
At the end of section {\bf IV} we'll discuss what happens in the full 
membrane theory beyond the self dual case.

Finally, in the remainder of this subsection, we will discuss  what is meant by taking the expectation value w.r.t the ``in'' states. Firstly, if one is taking expectation values in the ``in'' states one has to evaluate the $lim~r'\rightarrow -\infty$ of the ${\hat W}_2$ generator. The latter algebra has to be  expressed in terms of the quantum solutions of the continuous Toda theory. Secondly,  
setting aside for the moment the  singularities appearing in  the OPE of Toda field  exponentials, prior to taking the $r'=-\infty$ limit,  one has after evaluating (4.25) :

$$_{-\varphi(t)}<\rho |{\hat I}(t=2\pi)-{\hat I}(t=0)|\rho>_{-\varphi (t)}=
_{-\varphi(t=2\pi)}<\rho |{\hat I}(t=2\pi)|\rho>_{-\varphi (t=2\pi)}-$$
$$ _{-\varphi(t=0)}<\rho |{\hat I}(t=0)|\rho>_{-\varphi (t=0)}=I(2\pi)-I(0). \eqno (4.29)$$
after using the normalization condition $<\rho|\rho>=1$. Thirdly, modulo
the subtleties in pulling the expectation value inside the $t'$-integration,  
one can verify that the same  answer is obtained if one performs the expectation value before the $t'$ integration . One evaluates the action of an operator of $\rho$ acting on the ``in'' state as follows :

$${\cal F}[{\hat \rho}(r',t')] |\rho>_{-\varphi (t)}=lim_{r'\rightarrow -\infty,t'\rightarrow t}~{\cal F}[\rho (r',t')]|\rho>_{-\varphi (t)}. \eqno (4.30)$$ 
By inspection one can verify that after performing the integration one arrives at the same  (4.29) . This is the anlog of $f[{\hat x}]|x'>=f(x')|x'>$. The integral (4.24) clearly diverges as a result of the OPE of the Toda exponentials prior to taking  the $r'=-\infty$ limit. This can be fixed by introducing a ``normal ordering'' procedure that removes the short distance singularities as a result of the coincidence limit of two or more operators.  What is important is the $t$ integration. Since the l.h.s of (4.25) has a dependence on the $\lambda,{\bar \lambda}$ parameters the $\varphi (t)$ must also encode such a dependence. The spectral decomposition of $\varphi (t)$ can be taken to be :

$$\varphi_{\lambda, {\bar \lambda}} (t) =\sum_n A_n (\lambda,{\bar \lambda})sin(nt) +B_n (\lambda,{\bar \lambda})cos(nt). \eqno (4.31)$$
plugging (4.31) into (4.25) yields one equation with an explicit 
$\lambda,{\bar \lambda}$ dependence on both sides of the equation.
The other infinite number of integrals like (4.25) involving the other weights associated with the zero modes of the algebra generators are :

$${ \Delta}^{\lambda +\lambda^*}_k=  
_{-\varphi(t)}<\rho |[\int^{2\pi}_0 dt'~a_0
{\hat W}_s[{\hat \rho} (r'=-\infty,t')]]|\rho>_{-\varphi (t)}. \eqno (4.32)$$
with $k=2,3,4......\infty$ and $s=k+1 =3,4,.....\infty$ and the coefficient $a_0$ will be determined in ${\bf 4.3}$.
The infinite number of equations defines the ``coefficients'' appearing in the spectral decomposition of $\varphi (t)$, $A_n (\lambda, {\bar \lambda}),
B_n (\lambda, {\bar \lambda})$. 
The weights , after the dimensional reduction, are in general of the form :

$$\Delta_k ={1\over 2}\sum^k_{n=0} (a^k_n +a^{*k}_n)(\lambda +\lambda^*)^n. \eqno ( )$$  
.The weights are then  polynomials in $(\lambda +\lambda^*) $. Choosing the coefficients $ A_n= P_n (\lambda +\lambda^* )$, and 
$B_n= P_n (\lambda + \lambda ^*)$ to be polynomials of order $n$ in these variables  
will be sufficient to solve for (4.25,4.32). A real-valued polynomial of order $n$ has $n+1$ independent coefficients. Say one opts to choose : 
$A_n =a^n_o +a^n_1 (\lambda +\lambda^*)+.....+a^n_n (\lambda +\lambda^*)^n $. 
Each real-valued polynomial of order $n$ has $n+1$ independent coefficients, thus, the total number of independent coefficients in the infinite number of polynomials is of order 
$\sum_{n=o}^\infty (n+1)\sim n^2$, where one has  included  those coefficients stemming from the  $B_n$ terms as well. Now, the l.h.s of 
(     ) 

generates $k$ equations, 
each having  the weight-polynomials  of degree  $k+1$ in $\lambda$  containing $(k+2)$ known coefficients;  so the total number of known coefficients is $\sum_1^\infty (k+2)\sim k^2/2$. The rate of growth in the number of independent coefficients is the same in both cases. To have an exact match in the number of equations and in the number of independent variables one requires to impose a constraint on $A_n,B_n$ : $A_n=B_n$ and the doubling problem is eliminated. Although this is not strictly necessary since when $k\rightarrow \infty$ the number of equations equals the number of independent coefficients in the limit. 

Therefore, by matching, level by level in $k$, the coefficients involved in the powers of $(\lambda +\lambda^*)^0,(\lambda +\lambda^*)^1,(\lambda +\lambda^*)^2,,......$ one can solve for the $A_n=B_n$ polynomials  and determine the $\varphi (t)$ uniquely  in (4.31). 

Hence , an explicit  knowledge of the weights associated with a given representation fixes the form of the function $\varphi (t)$ and furnishes the $|\rho (in)>_{-\varphi (t)}$ state uniquely in terms of the $\lambda, {\bar \lambda}$ parameters. The same applies to the $|\rho (out)>_{\varphi (t)}$ state. Thus, the 
family of quantum states of  the putative quantum continuous Toda theory  associated with the highest weight representations can be  explicitly classified $after$ the dimensional reduction is taken place. There other states which correspond to other representations.

\bye

............................................

This is not the end of the calculation. We still have to go back to the original quantum version of eqs-(2.4) in order to have an exact quantization of the 
YM potentials that are in a $1-1$ correspondence with the membrane coordinates. We are assuming that the correspondence between the Toda theory and the YM theory also holds  off-shell. By  replacing Poisson brackets by commutators in (2.4), one can obtain the commutation relations of the YM operator-valued fields, ${\hat A}_y (\tau),{\hat A}_{{\bar y}}(\tau), {\hat A}_3 (\tau)$ in terms of the $\theta_l (\tau)$ fields. This was already discussed in ${\bf II}$  
 
The knowledge of the Hilbert space of states $|\rho_{\lambda,{\bar \lambda}} (in)>$ linked to the exact quantization of the continuous Toda field determines the Hilbert spaces associated with each one  of the infinite number of $\theta_l (\tau)$ fields that appear in the decomposition  : ${\cal H}_{\infty}={\cal H}_1\oplus 
{\cal H}_2 \oplus ......$. Hence,  
the spherical membrane in moving flat backgrounds ( the zero modes must be treated separately ) is an exact integrable quantum system insofar that it is equivalent to a $SU(\infty)$ YM theory dimensionally-reduced to one temporal dimension and the latter, in turn, can be re-expressed ( using the ansatz given by (2.4) ) in terms of an exact quantum integrable continuous Toda theory  ( with the  natural action of the  $U_{\infty}$ symmetry ). 

To conclude, the Hilbert space of states, $|\rho_{\lambda,{\bar \lambda}} (in)>$ can thus be  obtained ( in principle )  directly from eqs- (4.25,4.32) for any given highest weight irrepresentation of the $W_{\infty}$ algebra. This is the essence of this work. We must not forget that the symmetry algebra acting on the integrable system is the dimensionally- reduced $U_{\infty}$ algebra.  
....................................................................

x\smallskip
\centerline {\bf 4.3 The $U_\infty$ Algebra in the $Z,{\bar Z}$ Representation }
\smallskip
So far we have worked with integrals involving the $r=z+{\bar z},t$ variables. One could 
define the $new$ complex 
variables exploiting the fact that $t$ is a $periodic$ variable parametrizing a circle  [4] :

$$U\equiv t+ir.~{\bar U}\equiv t-ir.~Z=e^{-iU}=e^{r}e^{-it}.~{\bar Z}=e^re^{it} \eqno (4.33)$$
to obtain  a new complex domain $Z,{\bar Z}$ from the cylinder defined by the 
$U$ coordinates such that $r=-\infty$  corresponds to the 
origin $Z,{\bar Z}=0$ and  $r=+\infty$ corresponds to infinity. In this fashion a Hamiltonian analysis based on $r$ quantization ( the original time coordinate) corresponds to  a 
``radial'' quantization in the complex plane as  conventional CFT requires.  It is precisely when this correspondence is made that the two-dimensional world of non-critical $W_{\infty}$ strings matches the two effective dimensions which follow the dimensional reduction from the $2+1$ continuous  Toda theory to 
$1+1$ dimensions ; that of the $r,t$ variables. It is in this fashion how it makes sense to claim that non-critical $W_{\infty}$ strings in $D=27$ are connected to critical bosonic membranes in $D=27$ ( after the light-cone gauge is chosen so a $SU(\infty)$ YM theory dimensionally reduced to one temporal dimension  emerges ). Both theories are effectively two-dimensional. $W_{\infty}$ string theory amounts to an infinite collection of Virasoro-like strings with an infinite number of unusual central charges and intercepts ( linked to Regge-like trajectories ) [25]  . That infinite collection of strings effectively behaves like a membrane ( excluding the membrane's zero modes ). One should not worry at the moment about the fact that the area-preserving-diffs of the cylinder is $w_{1+{\infty}}$ algebra instead of the $w_\infty$ algebra associated with the plane. The presence of the $w_\infty$ structures orginate from the $3D$ continuous Toda theory. The physically relevant area!
 -preserving-diffs algebra which on
e is always referring to is the $su(\infty)$ algebra associated with the sphere.  

It is important to point out that the generator $W_2 [\rho (r,t)]$ (4.12) obtained as a dimensional reduction of (4.11) ( and its antichiral counterpart ) is a $mixed$ tensor w.r.t the $Z,{\bar Z}$ coordinates. This is easily derived by starting with  the integral of $W_2 [\rho (r,t)]$ w.r.t the ``angle'' variable $t$ to yield the energy. The  effective two dimensional space with coordinates, $r,t$, is a cylinder. $r$ plays the role of time and $t$ of space ( an angle ). Hence the conserved quantity w.r.t. the timelike Killing vector, $v^b = v^r(\partial/\partial r)$, is   
the energy given by the integral over a spacelike surface $\Sigma^a$ ( whose
oriented normal is timelike ) of the quantity : $T_{ab}v^b$ :

$${\cal H}=\int dt~W_2[\rho (r,t)]\equiv 
\int d\Sigma^aT_{ab}v^b =\int dt~T_{rr}  =   Energy. \eqno (4.34)$$ 
where $d\Sigma^r=dt;v^r=1$.
Because $T_{rr}(r,t)\equiv W_2 [\rho (r,t)]$ is one component of the stress energy tensor it must transform under the coordinate transformations : $r\rightarrow {1\over 2}ln(Z{\bar Z}).~t\rightarrow {1\over 2}ln(Z/{\bar Z})^i$ as follows :

$$T_{rr} (r,t)=W_2 [\rho (r,t)]={\partial Z\over \partial r}
{\partial Z\over \partial r}W_{ZZ} (Z,{\bar Z})+
{\partial {\bar Z}\over \partial r}{\partial {\bar Z}\over \partial r}
W_{{\bar Z}{\bar Z}} (Z,{\bar Z})+$$
$${\partial Z\over \partial r}
{\partial {\bar Z}\over \partial r}W_{Z{\bar Z}} (Z,{\bar Z})+
{\partial {\bar Z}\over \partial r}
{\partial Z\over \partial r}W_{{\bar Z}Z} (Z,{\bar Z}). \eqno (4.35)$$
An important remark is in order. Imagine knowing the remaining components 
$T_{rt},T_{tt},T_{tr}$ of the stress energy tensor in addition to $T_{rr}\equiv
 W_2[\rho (r,t)]$. In terms of the new complex coordinates a new tensor is constructed with components :$W_{ZZ};W_{{\bar Z}{\bar Z}}....$. Does this stress energy tensor generate new holomorphic/antiholomorphic transformations ? $Z\rightarrow Z'(Z);{\bar Z}\rightarrow {\bar Z}'({\bar Z})$ The answer is $no$. By inspection we can see, per example, that under dilatations  $z'(z)=\lambda z;{\bar z}'({\bar z})=\lambda {\bar z}$
 one does not generate $ Z'(Z)=\Lambda Z;{\bar Z}'({\bar Z})=\Lambda {\bar Z}$. And viceversa, in general $F(Z),{\tilde F}({\bar Z})$ does not always imply 
 $z'=f(z); {\bar z}'={\tilde f}({\bar z})$. Thus, the components of the new stress energy tensor cannot longer be split in terms of two $independent$  holomorphic/antiholomorphic pieces : $T(Z)=W^+_{(2,0)}(Z),{\tilde T}({\bar Z})=
W^-_{(0,2)}({\bar Z})$ as it occurs in ordinary CFT. The dimensional reduction $intertwines$ the holomorphic/antiholomorphic variables and the dimensional reduced 
$W_{\infty}\oplus {\bar W}_{\infty}$ algebra, is the symmetry  algebra $U_{\infty}$ acting on the Toda molecule,  not to be confused with the $V$ algebras constructed by Bilal. The $U_\infty$ algebra does $not$ generate conformal transformations in the $Z,{\bar Z}$ variables.

We can write down eqs-(3.9,4.15) again for the classical generator $W_2$   and by inspection  see that :

$$W_2[\rho (r,t)]\sim (\int^tdt'~\varphi (t'))^2=F(t)=F[(ln~(Z/{\bar Z})^i]\not= f(Z)+{\bar f}({\bar Z}). \eqno (4.36)$$
A splitting cannot occur unless a very special choice of $\varphi (t)$ is chosen . There is no reason why $F(U+{\bar U})=f(U)+{\bar f}({\bar U})$ unless one has a linear function. Another way of proving that there cannot be a split is by
computing :

$$<\lambda,{ \bar \lambda}| \oint {dZ\over 2\pi i}T(Z)|\lambda,{ \bar \lambda}>=
<L_{-1}>=0.~
<\lambda,{ \bar \lambda}| \oint {d{\bar Z}\over 2\pi i}{\tilde T}({\bar Z})|\lambda,{ \bar \lambda}>=
<{\tilde L}_{-1}>=0. \eqno (4.37)$$
where one has used the fact that the spurious sates 
$S=L_{-1}|\phi>,{\tilde S}={\tilde L}_{-1}|\phi>$ are orthogonal to the 
physical states, $|\phi>$. It is now fairly clear that the $<{\hat W}_2[\rho (r,t)]>$ cannot be expressed in terms  of the two expectation values given by (4.37). One cannot claim that the r.h.s of eq- (4.35) 
contains solely the $T(Z),{\tilde T}({\bar Z})$ pieces as ordinary CFT does.
The same argument applies to the terms in the r.h.s of (4.35) :$Z^2W_{ZZ},
 {\bar Z}^2 W_{{\bar Z}{\bar Z}}$. If one (wrongly) assumes that the mixed components are zero ( like in ordinary CFT) and that $W_{ZZ}$ is purely holomorphic and $W_{{\bar Z}{\bar Z}}$ is antiholomorphic,  after taking expectation values of the $L_1,{\tilde L}_1$ generators like it was performed in (4.37)  one gets zero. This wouldn't agree with $<W_2[\rho (r,t)]>$.
Shortly an explicit expression for the components of the r.h.s of (4.35) will be given.

Let us define a meromorphic real-valued field operator of two complex variables, $Z,{\bar Z},~{\hat W}_2 (Z,{\bar Z})$, to be determined in terms of $W_2[\rho (r,t)]$ admitting the following expansion in powers of $Z,{\bar Z}$ :

$${\hat W}_2 (Z,{\bar Z})=\sum_{N,{\bar N}}~{\hat W}^{N,{\bar N}}_2
Z^{-(n+im)-1}  {\bar Z}^{-(n-im)-1}. \eqno (4.38) $$
where $N\equiv n+im.~{\bar N}\equiv n-im$. The  mode expansion in the $Z,{\bar Z}$ variables (4.38) $differs$ from the expansion w.r.t the $z,{\bar z}$ variables as a result of the fact that the generator  $W_2[\rho (r,t)]=T_{rr} (r,t)$ is a particular component of a  $mixed$ stress energy tensor tensor  that intertwines $z,{\bar z}$. This is to be expected. 
Therefore, the correct expression to extract the residues is :

$${\cal P}[{ \Delta}^\lambda_1,{\bar  \Delta}^{\bar \lambda}_1] \leftrightarrow <\rho (in)|{\hat W}^{N=0,{\bar N}=0}_2 |\rho (in)>= 
<\rho (in)|[\oint{ dZ\over 2\pi i }\oint{d{\bar Z}\over
2\pi i}  {\hat W}_2 (Z,{\bar Z}) |\rho (in)>. \eqno (4.39)$$

If (4.39) is equal to (4.25) then certain conditions must be met . If we integrate (4.39) around circles of fixed-radius $R=e^r$ ( and then take the $R\rightarrow 0$ limit) one has after using : $dZ=-iZdt;d{\bar Z}=i{\bar Z}dt$ and taking into account an extra minus sign due to the counterclockwise/clockwise $Z,{\bar Z}$ contour integrations the following condition :

$$\int^{2\pi}_0{dt\over 2\pi}~\int^{2\pi}_0{dt\over 2\pi}~lim_{R\rightarrow 0}R^2{\hat W}_2 (Re^{-it},Re^{it})=\int^{2\pi}_0dt~a_0W_2[\rho (r=-\infty,t)]. \eqno (4.40)$$

hence one learns from (4.40) and (4.35) that :
$$a_0={1\over 2\pi}.~W_2[\rho (r,t)]=Z{\bar Z}{\hat W}_2 (Z,{\bar Z})=Z{\bar Z}(W_{Z{\bar Z}}+
W_{{\bar Z}Z}).~Z^2W_{ZZ}+{\bar Z}^2 W_{{\bar Z}{\bar Z}}=0. \eqno (4.41)$$

Later we shall study the conditions on the $f(t)=\sum a_n cos (nt) +b_n sin (nt)$ appearing in (4.24) 
in order to find the relations amongst the higher order modes than the zero ones and for all the $W_s$ generators in addition to the $W_2$.

Despite the fact that one has no longer purely  holomorphic/antiholomorphic transformations, the transformations generated by the $U_{\infty}$ generators  
${\hat W}_2 (Z,{\bar Z}) $ acting on ${\tilde \rho}(Z,{\bar Z})$ can, nevertheless,  be written as :

$$\delta^\epsilon_{W_s}~ {\tilde \rho}=\oint \oint {dZ\over 2\pi i} 
{d{\bar Z}\over 2\pi i }\epsilon (Z,{\bar Z}){\hat W}_s (Z,{\bar Z}){\tilde \rho }(Z',{\bar Z}'). \eqno (4.42)$$
after using the standard techniques of OPE and contour deformations of CFT. The contours sorround the $Z',{\bar Z}'$ variables and $not$ the origin after the contour deformation is performed. The $U_{\infty}$ algebra obtained from the dimensional-reduction process of the $W_{\infty}$ extended conformal field theory  inherits 
a similar transformation structure.

Going back to (4.39), the $|\rho (in)>_{-\varphi (t)}$ state at $r=-\infty$ is now the state corresponding to the operator insertion at the origin of the punctured  $Z,{\bar Z}$ plane.
To define the ``in'' state requires evaluating the limit :

$$|\lambda,{\bar \lambda}>\leftrightarrow lim_{Z,{\bar Z}\rightarrow 0}~
r(Z,{\bar Z}) {\hat \varphi}_{\lambda,{\bar \lambda}} [t(Z,{\bar Z})]~|0,0>. 
\eqno (4.43)$$.

The operator  quantity $-{\hat \varphi} (t)$ appearing in the quantum Toda solutions bears an explicit $\lambda,{\bar \lambda}$ depenence ( to be determined below ) and defines  the $|in>$ state at  $r=-\infty$ in  eq-(4.21). Now it makes sense to evaluate the contour integrals using the residue theorem  without obtaining a trivial zero answer.

If one opts to perform the contour integration $before$ the expectation value is taken in (4.39) is more convenient to take a contour surrounding the 
$origin$ which will absorb the singularities of the OPE in the coincidence limits $Z',{\bar Z}'\rightarrow Z,{\bar Z}\rightarrow 0$.  In general the contour integration does $not$ necessarily  commute with the evaluation process of taking the expectation value :
$$<W^{0,0}_2>\sim  \oint{ dZ\over 2\pi i }\oint{d{\bar Z}\over
2\pi i} <\rho_o|[ {\hat W}_2 ((Z,{\bar Z})] |\rho_o>. \eqno 
(4.44)$$
because one cannot naively  pull out the contour integrals outside the expectation values without introducing contact terms due to the time-ordering 
procedures ( in  the  expectation values ) leading to delta function singularities. Setting these subtleties aside, integrals (4.39,4.44) are  roughly the same.

If there were no singularities in the OPE defining the quantum ${\hat  W}_2(Z,{\bar Z})$ generator then (4.39) would be zero; per example, in the quantum case,  the expectation value of  : $<\rho_o|{\hat W}_2 |\rho_o>$ has singularities as result of  coincidence limits arising in the OPE of Toda exponentials  and for this reason integrals like (4.39) are not trivial. The singularities occur as $r'\rightarrow r\rightarrow -\infty$ which impies $Z'\rightarrow Z\rightarrow 0$ ( similarly with the 
${\bar Z}$ coordinates ) in the punctured complex plane. 
In the proceeding paragraphs we shall discuss the need to regularize certain expectation values   in order to evaluate  the conserved quantum integrals (charges )  of motion whose expectation values do not depend on the ``time'' variable; i.e. the quantum equations of motion for the operators obey : $d{\hat I}_n/dr =0;n=2,3....$ and these charges are involutive  $[{\hat I}_n, {\hat I}_m ]=0$.

How does one know that the right singularity structure appears in the evaluation of (4.39)? . To begin with the continous Toda field is $not$ a primary field  ( like the stress energy tensor in ordinary CFT)  and singularities of the required type do appear. In particular, the OPE of the exponentials of the ordinary Liouville field has been computed in great detail by Gervais and Schnittger [31] and the underlying quantum group structure was discovered associated with the algebra of chiral vertex operators. Savaliev has shown that a  realization of the $W_{\infty}\oplus {\bar W}_{\infty}$ algebra can be given $precisely$ in terms of the $3D$ continuous Toda field, and, $as~ such$,  one must have, accordingly, the correct singularity structure of the OPE in order to evaluate 
eq-(4.39). This is what a faithful realization of the $W_{\infty}$ entails.  Analogous to the results of [31],  
the OPE of the continuous-Toda fields exponentials should have singularities in the form of negative  
powers in $(Z-Z')^{-N}({\bar Z}-{\bar Z}')^{-{\bar N}}$ and the contour integrals
around the origin in  (4.39) can be computed yielding a non-zero result.

The same arguments can be applied to the $U_{\infty}$ algebra. The dimensional reduction shouldn't spoil ( in principle ) the presence of the singularities which appeared in the original $3D$ theory. One is free to work always in the $r,t$ representation where the equal time surfaces are the cylinders and $t$ is the angular variable. 
If one is worried about these technicalities one may quantize the classical $U_{\infty}$ algebra in the $r,t$ space by recurring to the explict quantum solutions (3.1.3.6,3.7) and the use of OPE based on the $r,t$ variables. If reparametrization invariance is not spoiled by the quantization process the results should be independent on which set of variables one decided to use. If the original $3D$ Toda theory is anomaly free the dimensional reduction process should be as well. The converse is not true. There is no reason why the supermembrane in $D=11$ is anomaly free just  because the $D=10$ superstring is.    

In (4.39) the states $|\rho (in)>$ are the sought-after highest weight states (ground states). The physical states belong to a particular class of the former. In order to obtain the space of physical states a complete list of all $unitary$ highest weight irrepresentations is required and from these a no-ghost theorem can be formulated which will select the restricted values of the infinite number of conformal weights
, $\Delta_k$, as well as selecting the value of the critical central charges ( or spacetime dimension as it occurs for the string ). This is tantamount of writing down the complete BRST cohomology 
of the physical vertex operators linked to the ( dimensionally-reduced ) $W_{\infty}$ CFT. This is a extremely arduous task because we don't have a $W_{\infty}$ CFT. Ordinary string theory is based on a $W_2$ CFT, rational or irrational CFT, and we know how difficult matters can be.  From each of these sought-after highest weight states
one builds a tower (the Verma module) by succesive applications of the ladder like operators (4.5). In ordinary string theory the tower of states are the so called spurious states. These are $physical$
when $D=26,a=1$ because the norm is zero. Eq-(4.39) is the analog of the string mass shell condition :$(L_o -a)|\phi>=0$ where $a=1$ results from the
zeta function regularization of the string zero-point-energy states. This is the reason why the expectation values of the zero mode operators $W_2^{N,{\bar N}}$ would require regularization as well. 

The computation of (4.39) is messy. Firstly, in order to compute the quantum analog of  eqs-(4.12) by replacing the classical continuous Toda solutions by the quantum operators,   would require to develop the OPE of Toda exponentials as well  as the  extension of the quantum enveloping algebra $U_q [sl(\infty)]$ associated with the vertex operator algebra; this hasn't been performed according to our knowledge.
This is the $major$ obstacle in obtaining the  exact quantization of the  membrane with the topology of a sphere. Once the OPE of the continuos Toda field exponentials is constructed
the tools to furnish  highest weight irreducible representations in terms of solutions of the quantum continuos Toda theory should be enough to quantize 
$exactly$ the spherical membrane; i.e. to build the fully fledged non-perturbative spectrum in apropriate ``$U_{\infty}$'' invariant  backgrounds.  The no-ghost theorem based on the unitary representations can then be studied to select critical spacetime dimensions. This is tentative ,of course. The main point is that the spherical membrane is an exact quantum integrable model and hence solvable. In the next section  simplified ways to solve this problem are  presented that allows to bypass the  explicit computation of the OPE. Thus, exact results can be  obtained. 

If the quantization of (4.12) displays singularities in the operator products so will the quantization of (4.15) and hence the energy is ill-defined. Therefore, a regularization is required so that the expectation value of (4.12) in the quantum case is finite. In ordinary CFT this is achieved  after a conformal normal ordering procedure ( or other suitable ordering ) is introduced so that correlations functions are finite and obey the Ward identities. The conformal normal ordering removes in most cases the infinities. It is customary to define the  normal ordered  product of two field-operators, $({\cal A}{\cal B})(w)$  as a contour integral surrounding $w$ of $(z-w)^{-1}{\cal A}(z){\cal B}(w)$. It is essentially a point-splitting regularization method.  The normal ordering does not obey the commutative nor associative property.         

With this lesson in mind let us introduce now the  family
of functions $\varphi_{\lambda,{\bar \lambda}} (t),d_{\lambda, {\bar \lambda}} (t)$ which appear in the general solution to the quantum continous Toda equation given by eqs-(3.1,3.6,3.7) ( to be determined below ) parametrized by $\lambda,{\bar \lambda}$; i.e to each 
highest weight irrepresentation parametrized in the form :$|\lambda,{\bar \lambda}>$
one associates the family of functions of $t$ parametrized by $\lambda,{\bar \lambda}$. We shall omit the suffixes in $\varphi (t)$ for the time being.
Due to the exact quantum integrability of the continuous Toda theory, the 
$<\rho (in)|{\hat W}_2|\rho (in)>$, after regularization, must have the same functional dependence  on  $\varphi (t)$ as the one given by the classical energy eq-(3.9). 
Inotherwords,  upon quantization a physical regularization prescription  must be one  such that the expectation value in the $|in>$  states   : 
 
$$  <\rho (in)|{\hat W}_2|\rho (in)>_{reg} = \int^{t_0}dt_1~ \varphi (t_1)
\int^{t_0}dt_1~\varphi (t_1). \eqno (4.45)$$
should agree with expression (3.9). The regularization method involves the point-splitting method discussed above. 
$\varphi (t)$ must now be interpreted as the expectation value ( to be determined shortly ) of the ${\hat \varphi} (t)$ operator which appears in the exact quantum solutions (given  by eqs-(3.1,3.6,3.7). 
These asymptotic expressions are given in terms of eq-(3.4) where now  $\varphi (t),ln~d(t)$ are operators obeying the commutation relations (3.8b). To simplify matters we choose $d(t)$ to be the unit operator so (3.8b)  constrains the $c$-number function $w(t)$ to satisfy 

$${1\over w(t)}{d^2 \over dt^2}  ({1\over w(t)})=0. \eqno (4.46)$$
so that eq-(3.7) has only an  $\hbar$ term in the delta function term.

The state $|\rho>$ in general 
 could be any state in the Hilbert space of states  associated with the operator ${\hat \rho}(r,t)$ for any value of $r$; i.e. it is the state associated with the $interacting$ highly nonlinear Toda field. At the moment we don't know if there is a $1-1$ correspondence between local fields inserted at particular locations in the punctured complex plane ( or other Riemann surface ) and states in this dimensionally-reduced $W_{\infty}$ CFT, as it happens with ordinary CFT. Choosing the asymptotic  states  simplify matters considerably  because the fields become free.  The asymptotic limit  was  uniquely determined in eq-(3.4) by the operator 
${\hat \rho}_o=(\partial^2 {\hat x}_o/\partial t^2)\rightarrow r{\hat \varphi} (t)$. 

A question comes to mind :``Where does  the $ \hbar$-dependence of the energy 
 ( given by the integral of eq-(4.15) w.r.t the  $t$ variable ) come from ?'' In eqs-(3.6,3.7)
the quantization was encoded, for the most part, in the function 
$w(t)$ which yields the $O( \hbar)$ corrections to the $\sum\varphi(t)^{-1}$ terms in eq-(3.1). In this  Heisenberg representation the quantum solutions for $\rho (r,t)$ are seen as operator valued quantities with an $explicit$ $\hbar$ dependence in (3.6,3.7). It is for this reason that $\varphi (t)$ must be  taken to be an operator ( as well as $d(t)$). After expectation values are taken in the asymptotic limits  the  explicit $\hbar$ factors appear in the eigenvalues of the operator ${\hat \varphi (t)}$ that  carry the $ \hbar$ dependence.  Whence the presence of the $ \hbar$. ( $d(t)$ was chosen  earlier to be the unit operator ). The eigenvalue equation needed to compute the expectation values in the 
$|\rho (in)>$ states  reads :

$$|\rho_{\lambda, {\bar \lambda}}> =lim _{Z,{\bar Z}\rightarrow 0} 
~r(Z,{\bar Z}){\hat \varphi}_{\lambda, {\bar \lambda}} [t(Z,{\bar Z})]~
|0,0>. \eqno (4.47)$$                  
The $ \hbar $ dependence is encoded in the eigenvalues/ weights. Per example, the angular momentum states  are $|J,J_z>$ such that 
${\hat J}_z |J,J_z>=m_z \hbar |J,J_z>$

Integrals like (4.39) are not the only ones required to determine 
$\varphi_{\lambda, {\bar \lambda}} (t)$
To find such an explicit $\lambda,{\bar \lambda}$ dependence of $\varphi (t)$ one needs to recur to $all$ the  weights of the representation, 
$  \Delta^\lambda_k$  ( and the
anti-chiral ones ). These  are also related to  the zero modes of Saveliev's realization of the chiral and antichiral $W_{\infty}$
algebras in terms of the dressed continous Toda field [4], where the infinite number of generators 
have similar  form  to eq-(4.11)  ( the number of  $t$ integrations depends on the value of the conformal spin )  : 

$${\cal W}^+_{(h,0)} [\partial \rho/\partial z;....\partial^h \rho /\partial z^h].
~{\cal W}^-_{(0,{\bar h})}
[\partial_z \rho \rightarrow \partial_{{\bar z}}\rho].~\partial {\cal W}^+/\partial {\bar z} =0.~\partial {\cal W}^-/\partial  z =0\eqno
(4.48) $$.  
After the dimensional reduction is taken, the expectation values of the zero modes,  expressed  in terms of the new variables, $Z,{\bar Z}$, of the $U_\infty$ generators    are :
$$<|{\hat W}^{N=0,{\bar N}=0}_s |>=
<\rho (in)|[\oint~{dZ\over 2\pi i}\oint {d{\bar Z}\over 2\pi i}   Z^{{s\over 2}-1}{\bar Z}^{{s\over 2}-1} {\hat W}_s (Z,{\bar Z})] |\rho (in)> 
\leftrightarrow {\cal P}[ \Delta^\lambda_k ,{\bar \Delta}^{\bar \lambda}_k]  . \eqno (4.49)$$
where $k\ge 1,~k=2,3,4......\infty$. The real-valued meromorphic field operator, ${\hat W}_s (Z,{\bar Z})$,  admits an expansion similar to (4.38) in powers of $Z^{-N-s/2}{\bar Z}^{-{\bar N}-s/2}$. 
We notice again a difference  between the exponents of $Z,{\bar Z}$ versus the usual ones in the $z,{\bar z}$ variables : $z^{h-1}{\cal W}^+_h;
{\bar z}^{{\bar h}-1}{\cal W}^-_{{\bar h}}$. 
The fact of having in some instances half-integer exponents in (4.49) is not that harmful; these are very natural in the fermionic string.

If eqs-(4.32) are to be equal to (4.49) similar conditions like (4.40) must be met. After performing the contour integrals around the origin by means of cirlcles of fixed radius  as it was done in (4.40) , 
using  $Z=Re^{-it},{\bar Z}=Re^{it},R=e^r$, yields :

$$ \int^{2\pi}_0{dt\over 2\pi }~lim_{R^2\rightarrow 0}(R^2)^{{s\over 2}}
 {\hat W}_s (Re^{-it},Re^{it}) =\int^{2\pi}_0dt~a_0
W_s[\rho (-\infty,t)]. \eqno (4.50)$$

this occurs for fixed radius, $R=e^r$, hence one learns that in general one must have  :

$$a_0={1\over 2\pi}.~(Z{\bar Z})^{s/2}{\hat W}_s (Z,{\bar Z})=W_s[\rho (r,t)]. \eqno (4.51)$$

The conditions on the other higher modes can be met also if the $n^{th}$ component, $f_n(t)$, of the function $f(t)$ obeys the following equation :

$$\int^{2\pi}_0dt~f_n(t)W_s[\rho (-\infty,t)] =\int^{2\pi}_0 {dt\over 2\pi}
lim_{R\rightarrow 0}~(R^{2})^{s/2}(R^2)^n (Z/{\bar Z})^{im}{\hat W}_s (Re^{-it},Re^{it}). \eqno (4.52)$$
where one $t$ integration absorbs a $2\pi$ factor.

An equality can be obtained iff $m={n\over 2} \Rightarrow N\equiv n(1+i/2);
{\bar N}\equiv n(1-i/2)$ and the $f(t)$ admits the expansion in $cosh~(nt); sinh ~(nt)$ instead of cosines/sines :
$f(t)=\sum_n a_ncosh~(nt)+b_nsinh~(nt)$. Plugging $f_n(t)$ in (4.52) yields after matching term by term in $n$ :

$$lim_{R\rightarrow 0}(R^2)^{s/2}{\hat W}_s(Re^{-it},Re^{it})=W_s[\rho (-\infty,t)]. $$ 
$$ a_ncosh~(nt)+b_nsinh~(nt)=lim_{R\rightarrow 0}~{R^{2n}e^{nt}\over 2\pi}. \eqno (4.53)$$
If the last relation holds at a $fixed$ value of $R=e^r$ and its generalization to other values of $Z,{\bar Z}$ this would require incorporating an $r$ dependence to the original $f(t)$ function so that the  $a_n (r),b_n(r)$ coefficients must  have an explicit $r$ dependence in order for (4.53) to be consistent.
The condition ( valid for other values of $Z,{\bar Z}$ ) is 

$$(Z{\bar Z})^{s/2}{\hat W}_s (Z,{\bar Z})=W_s[\rho (r,t)].~
a_n (r)={\tilde a}_n e^{2nr}.~b_n (r)={\tilde b}_n e^{2nr}. \eqno (4.54)$$
Eq-(4.53) yields :
$${\tilde a}_n \equiv lim_{R\rightarrow 0} {a_n (r)\over R^{2n}}. 
~{\tilde b}_n \equiv lim_{R\rightarrow 0} {b_n (r)\over R^{2n}}. \eqno (4.55)$$
so that the $R=e^r$ factors cancel out leaving  : 
$${\tilde a}_ncosh~nt+{\tilde b}_nsinh~nt={e^{nt}\over 2\pi}. \eqno (4.56)$$
in agreement with the previous conclusion that $a_0={\tilde a}_0 =(1/2\pi)$.

Concluding, we have that the real-valued meromorphic field 
${\hat W}_s (Z,{\bar Z})$ is related to $W_s[\rho (r,t)]={\tilde W}_s [{\tilde \rho}(Z,{\bar Z})]$ as follows :

$$(Z{\bar Z})^{s/2}{\hat W}_s (Z,{\bar Z})=W_s[\rho (r,t)]={\tilde W}_s [{\tilde \rho}(Z,{\bar Z})].\eqno (4.57a) $$  :
with the proviso that $f(t,r)$ admits the expansion with $r$ dependent coefficients shown above  and  :
$${\hat W}_s (Z,{\bar Z})=\sum_{N,{\bar N}}{\hat W}^{N,{\bar N}}_s Z^{-N-s/2}
{\bar Z}^{-{\bar N}-s/2}.~N=n+in/2;{\bar N}=n-in/2. \eqno (4.57b)$$
There is nothing wrong with the fact that the exponents of $Z,{\bar Z}$ in the  l.h.s of (4.57a) do not bear an $s$ dependence after premultiplying by the factor $(Z{\bar Z})^{s/2}$; the $s$ dependence is encoded in the components $ {\hat W}^{N,{\bar N}}_s$. (4.57a) is a realization of  the real-valued meromorphic field operators in terms of solutions of the continuous Toda molecue ; $\rho (r,t)$ after the change of variables is performed. It is clear from (4.57a) that the meromorphic fields are mixed and are no longer purely holomorphic nor antiholomorphic in the $Z,{\bar Z}$ variables.

Thus, 
Eqs-(4.39,4.49) 
are the equations we are  looking for.     
These are the eigenvalue equations which determines the very intricate relationship between 
$\varphi_{\lambda,{\bar \lambda}} [t(Z,{\bar Z})] $
and the weights $\Delta^\lambda_k$ ( and the antichiral ones). In order to find such a relationship one needs first to expand   :

$$\varphi_{\lambda, {\bar \lambda}}[t(Z,{\bar Z})] =
\sum_{m,{\bar m}}~A_{m{\bar m}} (\lambda, {\bar \lambda}) Z^{im} {\bar Z}^
{i{\bar m}} \eqno (4.58) $$
The reality condition on $\varphi (t)$ imposes the form of the exponents in the expansion. $(Z/{\bar Z})^{i}=e^{2t}$ is real. This fixes ${\bar m}=-m$ 
and convergence of the series implies certain restrictions on  
$A_{m,{\bar m}}$. Per example, the latter  cannot be all positive-valued functions if convergence occurs for  $m\rightarrow \infty$  or they must be in decreasing sequence. An expansion in terms of trigonometric functions of $t$ is also valid;  whereas   
 expanding in powers of $t$ is equivalent to expanding in powers of ${1\over 2}ln (Z/{\bar Z})^i$ which, in turn, can be Taylor expanded 
in powers of   :
$$ln (q+[(Z/{\bar Z})^i-q])=lnq+ln (1+[{(Z/{\bar Z})^i\over q}-1])=
lnq +[{(Z/{\bar Z})^i\over q}-1]-{1\over 2}[{(Z/{\bar Z})^i\over q}-1]^2 +....
\eqno (4.59)$$
where $q$ is chosen to ensure convergence of the Taylor series; i.e. the $ln (1+x)$ converges for $|x|<1.$ The variable $t$ is an ``angle'' variable  ranging from $[0,2\pi]$. The coefficients,  $A_{m{\bar m}}$, are  functions ( to be determined )  of the $\lambda, {\bar \lambda}$ parameters
characterizing the representations (like the weights). This is precisely where a regularization of the expectation values of the integrals 
(4.39,4.49) 
is required in the same vain that the expectation $<L_0>=a$ required a zeta function regularization for the string. To define the $|\rho (in)>$ state requires the evaluation of the $r=-\infty$ limit of the quantity  $r(-\varphi (t)$ which appears in the asymptotic regime in the solutions like 
(3.1) 
and, as specified in 
(4.21), 
requires a change in sign to assure convergence at $r=-\infty$. To define the ``in'' state requires to evaluate the limit :

$$lim_{Z,{\bar Z}\rightarrow 0}~ln(Z{\bar Z})
\sum_{m,{\bar m}}~A_{m{\bar m}} (\lambda, {\bar \lambda}) Z^{im} {\bar Z}^
{i{\bar m}} \eqno (4.60) $$
The powers $(Z/{\bar Z})^{im}$ are not well defined ( although finite, $e^{2mt}$) in the $Z,{\bar Z}\rightarrow 0$
limit. the regularization prescription must be for all values of $m$  :
$$ lim_{Z,{\bar Z}\rightarrow 0}~ln(Z{\bar Z})
A_{m,- m} (\lambda, {\bar \lambda})Z^{im}{\bar Z}^{-im}= 
A^{reg}_{m,-m} (\lambda, {\bar \lambda}) e^{2mt}.\eqno (4.61)$$
The presence of $t$ in 
(4.61) 
is  due to the ambiguity of the zero limit  and is important. From 
eqs-(4.25,4.29) 
one learns that  the ``in'' state, $|\rho_{\lambda,{\bar \lambda}}>$   bears an intrinsic $t$ dependence encoded in the function $\varphi (t)$. After computing the inner products 
; $ <\rho|\rho>=1$ this $t$ dependence drops out appearing only on the limits 
$t=0,t=2\pi$ as shown explicitly in 
(4.29). 
In this fashion the ambiguity is removed.   Therefore, choosing the infinite number of  ``coefficients'' to absorb the logarithmic singularity regularizes  the expectation values of the integrals . Per example, after the contour integrations absorb the singularities in the OPE of the Toda exponentials, the regularized expectation value of the zero modes of the $W_2(Z,{\bar Z})$ operator/generator is  :

$$<\rho_{\lambda,{\bar \lambda}}|{\hat W}_2^{N=0,{\bar N}=0}|\rho_{\lambda,{\bar \lambda}}>_{reg} ={\cal F}_2[A^{reg}_{m,{\bar m}} (\lambda,{\bar \lambda })]={\cal P}[  \Delta^\lambda_1, {\bar \Delta}^{\bar \lambda}_1 ]. \eqno (4.62)$$
and similarily for $s=3,4,5,......\infty$
$$<\rho_{\lambda,{\bar \lambda}}|W^{N=0,{\bar N}=0}_s|\rho_{\lambda,{\bar \lambda}}>_{reg} ={\cal F}_s[A^{reg}_{m,{\bar m}}(\lambda, {\bar \lambda})]=  
{\cal P} [\Delta^\lambda_k ,{\bar \Delta}^{\bar \lambda}_k.] \eqno (4.63)$$
The r.h.s of 
(4.62,4.63) 
is given by the prescription in 
(4.26b)).
The infinite number of eqs 
(4.62,4.63) 
are the ones obtained after the evaluation of 
(4.39,4.49) 
using the regularization prescription 
(4.61) for the states as well as the point-splitting for the operators involved in the definition of the $W_s$ generators.
The ${\cal F}_s;$ for $s=2,3,4.....\infty$ are an infinite number of  known functionals of the regularized ``coefficients `` $A^{reg}_{m,{\bar m}}$ that bear the sought-after $\lambda,{\bar \lambda}$ dependence encoded in the $\varphi (t)$ function given by 
(4.58)
Having an infinite family of functions in 
$\lambda, {\bar \lambda},~{\tilde \Delta}^\lambda_k,...~k=1,2......$, the integral equations 
(4.62,4.63) 
for $s=2,3,4........$,
will be sufficient ( in principle ) to specify  $A^{reg}_{m{\bar m}}(\lambda, {\bar \lambda});~m=0,1,2......$ enabling to establish the $|\lambda,{\bar \lambda}>\rightarrow |\rho_{\lambda{\bar \lambda}}>$ correspondence. Therefore,  
these  infinite number of equations, would allow us  to construct 
the states $|\rho_{\lambda {\bar \lambda}}>$ and quantize the theory exactly.  
\smallskip
\centerline{\bf 4.4 CASIMIRS}
\smallskip

One way to obtain explicit answers without having to compute the OPE of Toda 
exponentials is by recurring to the construction of the infinite number of Casimirs. In ${\bf V}$ other ways to obtain exact answers are found. The 
infinite number of involutive  $regularized$ quantum integrals of motion, the Casimirs, are    [4] :

$$I_n [(<\rho|{\hat \varphi}_{\lambda,{\bar \lambda}} (t)|\rho>_{reg})] =<\rho|{\hat I_n}|\rho>_{reg} \sim \int^{2\pi}_0~dt~(\int^t~dt' 
\varphi_{\lambda,{\bar \lambda}} (t'))^n.
 \eqno (4.64)$$

Having an explicit solution for $\varphi_{\lambda,{\bar \lambda}}(t)$ automatically yields the expectation values of the infinite number of Casimirs of the $U_{\infty}$ algebra. The explicit expression relating the infinite number of involutive conserved charges in terms of the generators of the chiral $W_{\infty}$ algebra has been given by Wu and Yu [33] :

$${\hat Q}_2=\oint~{\hat W}_2 dz;~{\hat Q}_3 =\oint~{\hat W}_3 dz.~
{\hat Q}_4 =\oint~({\hat W}_4-{\hat W}_2.{\hat W}_2)(z)dz.$$

$${\hat Q}_5 =\oint~({\hat W}_5-6 {\hat W}_2.{\hat W}_3)(z)dz;
{\hat Q}_6 =\oint~({\hat W}_6  -12 {\hat W}_2.{\hat W}_4 -12 {\hat W}_3 {\hat W}_3 +8 {\hat W}_2 {\hat W}_2{\hat W}_2
                            )(z)dz;....\eqno (4.65)$$

Similar expressions hold for the antichiral algebra. In the dimensionally-reduced $U_{\infty}$ algebra case, these expressions hold as well, where now the integrals to use are those of the type outlined in 
eqs-(4.24) 
which  replace the contour integrals in the complex plane $z,{\bar z}$ by integrals w.r.t. the $t$ variable.  The $f(t,r=-\infty)$ terms  can be set to a constant ( the zero mode is selected ). Upon evaluation of expectation values and 
a regularization  one will have then expressions of the type :
$$
<{\hat I}_2>_{reg}=E= <\int dt ~{\hat W}_2 [\rho (r=-\infty,t)]>_{reg};
~<{\hat I}_3>_{reg} = <\int dt ~{\hat W}_3 [\rho (-\infty,t)]>_{reg}.$$
$$ <{\hat I}_4>_{reg} =<\int dt ~({\hat W}_4-{\hat W}_2.{\hat W}_2)
[\rho(r=-\infty,t)]>_{reg}.
....\eqno (4.66)$$

The values of the expectation values in (4.66) should  agree with the Casimirs (4.64). These expressions could have been given in terms of the $Z,{\bar Z}$ as well. 
A question immediately arises :
How does one know that a removal of infinities by a suitable regularization procedure yields precisely eqs-(4.64,4.66) for the expectation values ? The answer lies in the $complete$ quantum integrabilty property of the quantum continuous Toda theory. If the theory is integrable the expectation values of the Casimirs ( after regularization effects) should agree with the first term of (4.64). 
If one knows $a~ priori$ the explicit $\lambda, {\bar \lambda}$ dependence of the Casimirs, a dimensional reduction yields the Casimirs of the $U_\infty$ algebra and eqs-(4.64) yield automatically the required relations sufficient to determine  the  $\lambda,{\bar \lambda }$ dependence of $\varphi (t)$ $without$ having to evaluate the OPE in 
eqs-(4.25,4.32).
 Therefore, a prior knowledge of the expectation values of the Casimir operators  of the $W_\infty$ algebras would furnish the correspondence between $|\lambda, {\bar \lambda}>\rightarrow |\varphi_{\lambda, {\bar \lambda}} (t)>$ without the need to evaluate the OPE.    

The infinite number of eqs-(4.25,4.32) 
are almost impossible to solve at the moment for the reason that we don't have the OPE rules for the exponentials of continuous Toda fields. Unless an explicit construction of the Casimirs associated with $W_{\infty},U_{\infty}$ algebras is known.  
As far as we know the Casimirs  have not been constructed. There are  special cases when we can  have exact solutions. This is discussed  in the following section.

To summarize : given a quasi-finite highest-weight irreducible representation of the $W_{\infty},{\bar W}_{\infty}$ algebras; i.e. given the generating function for the infinite number of conformal chiral ( antichiral ) weights
: ${\tilde \Delta}^\lambda (x)\Rightarrow \Delta^\lambda_k$ and the central charge ; $C $ and $b(w),\chi (characters)....$ one can ( in principle ) from 
eqs-(4.25,4.32) 
determine $\varphi (t)$
as a family of functions parametrized by $\lambda,{\bar \lambda}$.  
Since the latter are continuous parameters the energy spectrum 
(3.9,4.15)
is
$continuous$ in general. One has a continuum of highest weight states.  Below we will study a simple
case when one has a discrete spectrum characterized by the positive integers $n\ge 0$.    
Once $\varphi_{\lambda{\bar \lambda}} (t)$ is determined  the  regularized Hamiltonian operator  obeys the  equation :

$${\hat H} ~|\varphi_{\lambda {\bar \lambda}} > = E [\varphi_{\lambda{\bar \lambda}} (t)] ~ |\varphi_{\lambda {\bar \lambda}} >.       
\eqno (4.67)$$
where the Energy eigenvalue is (3.19). An explicit knowledge of 
$\varphi (t)$ can be given once 
eqs-(4.25,4.32) 
are solved.
Instead of having quantum numbers say  $|n,l,m_l...>$ like  in the hydrogen atom, per example, here  one has for ``eigenvalues'' integrals of suitable functions of $t$ which are the regularized expectation values 
of expressions involving the operator ${\hat \rho }(r,t)$. To know $\varphi (t)$ requires finding out  the values of the infinite number of coefficients, $A_n,B_n$ in 
(4.31) 
that play the role of the  ``spectra''. This is a very difficult task.  We proceed in ${\bf V}$ to find particular exact solutions.

\centerline{\bf V. Discrete Spectrum}

Below we will study a simple
case when one has a discrete spectrum characterized by the positive integers $n\ge 0$; i.e. the $|\lambda>=|n>$. We shall restore now the coupling
$\beta^2<0$ given in (2.13) . A simple fact which allows for the possibility of discrete energy states is to use the
analogy of the Bohr-Sommerfield quantization condition for periodic system. It occurs  if one opts to choose for the quantity 
$exp[\beta \varphi (t) r]\equiv exp[i\Omega r]$ which appears in (2.14); $\Omega$ is the frecuency parameter ( a constant ). For this new choice of $\varphi (t)$ the expression (3.9) for the classical energy needs to be modified in general. Below we will show that if $d(t)$ is chosen to be zero then (3.9) is still valid. When 
the dynamical system is periodic in the variable $r$ 
with periods $2\pi /\Omega$, a way to quantize the values of $\Omega$ in units of $n$ is to recur to the Bohr-Sommerfield
quantization condition for a periodic orbit :
$$ J=\oint~pdq =n\hbar       \eqno (5.1)$$ 

which reflects the fact that upon emission of a quanta of energy $\hbar \Omega$ the change in the enery level as a function of
$n$ is  [34]  :

$$  \partial E/\partial n =\hbar \Omega ={2\over 3} (2\pi)^3 (\hbar )^2\Omega \partial \Omega /\partial n.\Rightarrow $$

$$\Omega (n) ={3\over 2 (2\pi)^3 }{ n\over \hbar}. \eqno (5.2)$$
Hence the energy is 

$$E={3\over 4}(2\pi)^{-3} n^2.    \eqno (5.3)$$
which is reminiscent of the rotational energy levels $E\sim h^2l(l+1)$ of a rotor in terms of the angular momentum quantum
numbers $l=0,1,....$. In order to have a proper match of dimensions it is required to insert the membrane tension as it happens
with the string. In order to classify the physical set of states we have to have at our disposal of all of the unitary 
highest weight irrepresentations. As far as we know these have not been constructed for $W_{\infty}$. For $W_{1+\infty}$ these
have been constructed by Kac and Radul [7] and by the group [6]. These representations turned out to have a crucial importance 
in the classification of some of the Quantum Hall-Fluid states [11].

Saveliev [4] chose the $\varphi (t)$ in (3.1) to be negative real functions to assure that the
potential term in the Hamiltonian vanished at $r\rightarrow \infty$ and arrived at (3.9). 
In  case that the functions $\varphi (t)$ are no longer $<0$; i.e when $\beta \varphi r$ is no longer a real valued quantity
$<0$,  the asymptotic formula (3.9) will no longer hold and one will be forced to perform the very complicated integral !

$${\cal H} = \int~dt[-{1\over 2}\beta^2(\partial p/\partial t)^2 +({m^2\over \beta^2}) exp~[\beta\partial^2x/\partial t^2 ]
]. \eqno (5.4)$$

where $p= \beta \partial x/\partial r$ is the generalized momentum corresponding to $\rho \equiv \beta\partial^2x/\partial
t^2,$ and 
 $\mu^2 \equiv ({m^2\over \beta^2})$ is the perturbation theory expansion parameter discussed in [5]. Without loss of
generality it can be set to one. Nevertheless, eqs-(5.1,5.2) are still valid. One just needs to evaluate the
Hamiltonian at $\Omega r=2\pi p$ where $p$ is a very large integer $p\rightarrow \infty$ and take $d(t)=0$ in (3.2,3.3) ( the logarithm is illdefined, nevertheless the energy is still finite ) :

$$exp[\partial^2 x/\partial t^2]\rightarrow d(t)exp[i2\pi p] =0.~(\partial p/\partial t)^2\rightarrow (\int \varphi
dt')^2...\eqno (5.5)$$   recovering (3.9) once again.  

Are there zero energy solutions ?. If one naively set $\varphi (t) \equiv 0$ in (3.2) or set $n=0$ in (5.2) one would get a zero
classical energy. However eqs-(3.2,3.3) for the most part  will be singular and this would be unacceptable. One way  zero
energy states could be obtained is by choosing $\varphi (t), d(t)$ appropriately so that (5.4) is zero. Since one has one
equation and two functions to vary pressumably there should be an infinite number of solutions of zero energy. In the quantum case one has an extra function
to deal with $w(t)$ so it is possible to set $\varphi $ to zero and use $d(t),w(t)$ appropriately to avoid singularities. 

At first sight there could be an infinity of quantum ground states. In the discrete spectrum case the lowest of the ground states ( the lowest in energy of 
the highest weight states) corresponds to $n=0$. From this state one then builds a representation by erecting the tower of states (4.5). 
As mentioned earlier we do not know whether the discrete states are the physical ones. The no-ghost theorem has yet to be constructed.
For this reason it is of paramount importance to have a list of all unitary highest weight irreps in order to avoid negative norm
states. In the string picture one has that the central charge must be $26$ and the Regge intercept must be $a=1$. The membrane, as a 
non-critical $W_{\infty}$ string, is comprised of an infinite set of Virasoro type of strings with unusual central charges and intercepts
[25]. Therefore, we expect to have a selected value for the conformal weights $\Delta_k$ as well as the central charge.

Solving (4.67) is analogous to solving a time independent  ``Schroedinger''-like equation. Concentrating on the case that $\varphi (t) <0 $;
 the wave functional is defined :
$\Psi [\rho,t]\equiv <\rho|\Psi>$ where the state $|\rho>_{\varphi}$ has an explicit dependence on $\varphi$ which also depends
on $\lambda$ as shown in section ${\bf IV}$. Upon replacing 

$\partial p/\partial t \rightarrow -i\hbar (\partial/\partial t \delta/\delta \rho)$
as an operator acting on the $\Psi$, the time-independent equation for the wave functional becomes :

$$\int^{2\pi}_0 dt~[(-i\hbar \partial/\partial t \delta/\delta \rho)^2 +exp~\rho ] \Psi [\rho (r't');t] = 
\int^{2\pi}_0 dt~(\int^t dt'\varphi (t'))^2 \Psi [\rho (r',t');t].
\eqno (5.6)  $$

One could have written (5.6) in the momentum representation :$ \rho \rightarrow -i\hbar \delta/\delta p$ acting on the "Fourier" transfom
of $\Psi$.

The action functional is :

$$\int dt\int~{\cal D}\rho dr \Psi^+(i\hbar {\partial \over \partial r} -{\cal H})\Psi [\rho (r',t');r,t]. \eqno (5.7)$$

${\cal D}\rho$ is the functional integration measure; $r$ is the variable linked with the physical time
and the on-shell condition is just (5.6).
This is the second-quantization of the physical quantities. $\rho (r,t)$ has already been first-quantized in (3.6,3.7). 
One must  $not$ interpret $\Psi$ as a probabilty amplitude but as a field operator  which creates a
continuous Toda field in a given quantum state $|\rho>_{\varphi}$ associated with the classical configuration configuration given
by eq-(3.1) in terms of $\varphi_{\lambda} (t)$. The functional differential equation (5.6) is extremely complicated.  
A naive zeroth-order simplification will be given shortly. This is  because the $\Psi$ can have the form $\Psi =\Psi [\rho,\rho_{t'},\rho_{t't'},.......]$.                                                                  
The  equation in the momentum representation does not have that complexity but it has an exponential functional differential operator.
Evenfurther, the $\Psi$ is a non-local object. This is of no concern : In string field theory the string field is a multilocal object that depends on 
all of the infinite points along the string.

One can expand $|\Psi>$ in an infinite dimensional basis spanned by the Verma module  (4.5) associated with the state
$|\lambda>$ and let $\lambda$ run as well over all the highest weights. Given a vector $v_\lambda \epsilon  {\cal V_\lambda}$ ( Verma module)  one has :

$$|\Psi > = \sum_{\lambda}\sum_{v_\lambda}~<v_\lambda ||\Psi > |v_\lambda >.~v_\lambda \epsilon {\cal V_\lambda}    \eqno (5.8)$$

This is very reminiscent of the string-field $\Phi [X (\sigma )] =<X||\Phi (x_o)>$ where $x_o$ is the center of mass coordinate
of the string and the state $|\Phi (x_o)>$ is comprised of an infinite array of point fields :

$$|\Phi (x_o)> =\phi (x_o) |0> +A_\mu (x_o) a^{\mu +}_1 |0> +g_{\mu\nu}a^{\mu +}_1 a^{\nu +}_1 |0> +.....\eqno (5.9)$$

where the first field is the tachyon, the second is the massless Maxwell, the third is the massive graviton....In the  string
case one does not customary expand over the towers of the Verma module since these states  have zero  norm. However one should include
all of the states. The oscillators play the role of ladder-like operators acting on the ''vacuum''$|0>$ in the same manner that the Verma module is
generated from the highest weight state $|\lambda>$ by succesive application of a string of $W(z^{-n}D^k)$ operators acting on
$|\lambda>$.
The state $|\rho (r,t)>_{\varphi}$ 
It is the relative of the string state $|X(\sigma_1,\sigma_2)>$ whereas $|\Psi >$ is the relative of the string field state 
$|\Phi >$. The naive zeroth order aproximation  of the ``Schroedinger''-like equation is of the form :

$$[\partial_t^2 \partial_y^2 +e^y] \Psi (y,t) =E\Psi (y,t). \eqno. (5.10)$$

A change of variables :$x=2e^{y/2}$ converts (5.10) into  Bessel's equation after one sets $\Psi (y,t) =e^t\Phi (y)$ or
equal to $e^{-t}\Phi (y)$  :   

$$  (x^2\partial^2_{x^2} +x\partial_x +x^2-4E)\Phi (x)=0. \eqno (5.11)$$

and whose solution is :$\Phi (x) =c_1{\cal J}_\nu (x) +c_2 {\cal J}_{-\nu} (x)$
where $\nu \equiv 2\sqrt E$ and $c_1,c_2$ constants.

The wavefunctional is then a linear combination of  :

$$\Psi [\rho(r',t');t] = e^t\int\int~dr'dt'[c_1{\cal J}_\nu (2e^{\rho (r',t')/2}) +c_2{\cal J}_{-\nu} (2e^{\rho (r',t')/2})]  \eqno (5.12)$$
or the other solution involving $e^{-t}...$.

One may notice that  discrete energy level (in suitable units such as  $\nu=2\sqrt E=n$) solutions are possible. Earlier
we saw in (5.2) that $E(n)\sim n^2$ so $\sqrt E \sim n$. Therefore setting $2\alpha e^{y/2\alpha}=x$ where $\alpha$ is a 
suitable constant allows to readjust $\nu =2{\sqrt {\alpha E}}=n$. The
Bessel functions will have nodes at very specific points . The solutions in this case will be given in terms of ${\cal J}_n$ and
the modified Bessel function of the second kind, $K_n$. These solutions are tightly connected with the boundary conditions of
the wave-functional.

The other instance in which exact solutions can be obtained is when the OPE of the Toda exponentials is available. There exists a particular class of solutions to the $3D$ continuos Toda field equation that bears a direct relationship to the Liouville equations :
$${\partial^2 \over \partial z \partial {\bar z} }u (z,{\bar z},t)={\partial^2 e^u \over  \partial t^2}. \eqno (5.13)$$

The ansatz [4] 

$$e^{u(z,{\bar z},t)} =(\alpha t^2 +\beta t+\gamma)e^{\phi (z,{\bar z})}\Rightarrow 
~\partial_z \partial_{\bar z} \phi (z,{\bar z})=e^{\phi (z,{\bar z})}             \eqno (5.14)         $$
so a particular class of solutions of (5.13) in terms of those solutions of the Liouville equation (5.14) can be  found . The solutions of (5.13) are tightly connected to 
Killing symmetry reductions of Plebanski heavenly equations for $4D$ self dual gravitational  backgrounds. Evenfurther, these hyperkahler/heavenly  backgrounds exhibit duality symmetries via 
the Legendre transform. For a recent discussion on the status of   
non-abelian  duality symmetries associated with  the non-linear supersymmetric $\sigma$ models and $WZNW$ models relevant to  superstring theory, see 
Sfetsos [42].  
In [14] we have recently  derived a straightforward  origin of the analog of $S,T$ duality symmetries for all $p$ branes using  a reformulation of extended objects as composite antisymmetric tensor field theories of the volume-preserving diffeomorphism group  based on the work of [13]. The upshot of this formulation is that $S,T$ duality emerge in a very natural way from the start. The quantization program remains yet to be done. 

Thus, for the special class of solutions (5.13) one can compute the OPE of the 
Toda exponentials in terms of Liouville field theory  following the 
fusion rules constructed by Bilal,Gervais and Schnittger [32]. Quantum $tau$ functions can also be explicitly built [40].   
For this particular case we can have also  explicit results of  ${\bf IV}$.

\centerline{\bf VI. CONCLUSION AND CONCLUDING REMARKS}

An exact integrable set of quantized solutions of the spherical membrane moving in flat target backgrounds has been obtained. Such  a special class of solutions are those related to dimensional reductions of $SU(\infty)$ YM theories from ten to four 
dimensions. The latter are constructed in terms of instanton solutions in $D=4$ that can be related to the continuous Toda theory after an ansatz is made. Not surprisingly, the system is integrable. We don't know the fate of more general class of solutions of the YM equations nor what happens for surfaces whose topology is not spherical nor the case of membranes moving in curved backgrounds. This deserves further study. We hope to have advanced the need to built unitary highest weights irreps of $W_\infty$ algebras and the construction of the OPE of continuous Toda exponentials required to furnish the full non-perturbative membrane's spectrum.  

What will happen with other $p-branes/extendonds$ ? A reformulation of $p$ branes has been recently been  constructed by the author [14] based on [13]. Extended objects can be seen as new $composite$ antisymmetric tensor field theories of the volume-preserving diffs group. 
The positive side to this reformulation is not only that the volume-preserving diffs symmetry is  manifest but $S,T$ duality emerge in a very natural way from the very beginning. The next step is to find explicit classical solutions 
and use the representation theory of the volume-preserving-diffs to construct the spectrum. 
Membranes appear in a wide variety of physical models .

1-.It has been argued that a four dimensional 
anti-deSitter spacetime, $AdS_4$, whose boundary is  $S^{2}\times S^1$, could be realized as a membrane at the end of the
universe. In particular, singleton field theory can be described on the boundary of $AdS_4$ where singletons are the most
fundamental representations of the de Sitter groups [36]. Moreover, on purely kinematical grounds, infinitely many massless
states of all spins (massless in the anti-de-Sitter sense) can be constructed out of just two singletons ( preons). In
particular, the $d=4~N=8$ Supersingleton field theory formulated on the boundary of $AdS_4$ bears a connection with the
supermembrane moving on  $AdS_4\times S^7$ . The rigid $OSp(8|4)$ symmetry acts as the superconformal group on the boundary 
$S^{2}\times S^1$ [36].  In view of this it is important to study if there is any connection between the wave-functional
behaviour of ${\bf V}$ at the boundaries and singleton field theory. The supersymmetric Toda equation has been discussed by [5], thus in principle the results found here could be extended to the supersymmetric case.

Roughly speaking, a membrane is comprised of an infinite number of strings. Thus the membrane can be seen as a coherent state of
an infinite number of strings . This is reminiscent of the Sine-Gordon soliton being the fundamental fermion of the massive
Thirring model, a quantum lump [34]. The lowest fermion-antifermion bound state (soliton-antisoliton doublet) is the fundamental
meson of Sine-Gordon theory. Higher level states are built from excitations of the former in the same way that infinitely many
massless states can be built from just two singletons.

2-.Edge states : Recently [11], the set of unitary highest weight irreps of $W_{1+\infty}$ have been used to algebraically characterize the low 
energy edge-excitations of the incompressible ( area preserving) Quantum Hall Fluids. Experimental data matching the Jain hierarchy
for certaing filling factors was identified with certain minimal models. A surprising feature was found : Non Abelian structures
were discovered and neutral quark-like excitations with $SU(m)$ quantum numbers and fractional statistics as well. This merits
a further investigation from the membrane picture : a dimensionally reduced $SU(\infty)$ Yang-Mills theory.                        

3-.Noninteracting multi string solutions in curved spacetimes were studied in 
[37]. A single world sheet simultaneously 
decsribed many different and independent strings with no analog in flat space which appears as a consequence of the coupling
of the strings with the curved spacetime (de Sitter). This ``multistring'' picture fits with the membrane picture although one has to look at the propagation in 
anti de Sitter backgrounds.

4-.Perhaps the most relevant physical applications of the membrane quantization program will be in the behaviour of black hole
horizons [8]. These have been described in terms of a dynamical surface whose quantum dynamics is precisely that of a
relativistic membrane. Thermodynamical properties like the entropy and temperature of the black hole were derived in agreement
with the standard results. Results for the level structure of black holes were given. A ''principal'' series of levels was found
corresponding to the quantization of the area of the horizon in units of the ''area quantum'' :$A=nA_o.~A_o =8\pi$. From each
level of this principal series starts a quasi-continuum of levels due to the membrane's excitations. The connection between black hole physics and non-abelian Toda theory has been studied in [9]. $W$ gravity was formulated as chiral embeddings of a Rieman surface into $CP^n$. Toda theory plays a crucial role as well [10].

5-. The ordinary bosonic string has been found to be a special vacua of the $N=1 $ superstring [12]. It
appears that there is a whole hierarchy of string theories : $w_2$ string is a particular vacua of the $w_3$ string and so
forth......If this is indeed correct one has then that the (super) membrane, viewed  as  noncritical $W_{\infty}$ string theory,
is, in this sense, the universal space of string theory. 
The fact, advocated by many, that a Higgs symmetry-breakdown-mechanism of the infinite number
of massless states of the membrane generates the infinite massive string spectrum fits within this description.

6. Finally, we hope that the essential role that Self Dual $SU(\infty)$ Yang-Mills theory has played in the origins of the
membrane-Toda theory, will shed more light into the origin of duality in string theory [38,39]. For a review of duality in string theory and the status of string solitons see [40]. An important review of extended conformal field theories
 see [43].

As of now  we must have all unitary irreps of $W_{\infty}$ and be able to construct the OPE of the Toda exponentials to  fully exploit the results of ${\bf IV}$. The discrete spectrum solution warrants a further investigation and the supersymmetric sector as well.

\smallskip

ACKNOWLEDGEMENTS. We thank M.V. Saveliev for many helpful suggestions concerning the exact quantization program of the continuos Toda theory. 
To G.Sudarshan, Y.Ne'eman, C.Ordonez, J.Pecina, B. Murray for discussions  and to the Towne family for their kind and
warm hospitality in Austin, Texas. This work was supported in part by a ICSC, World Laboratory Fellowhip.

 \smallskip

\centerline {\bf REFERENCES}

\bye